\documentclass[reprint,amsmath,amssymb,aps,prb,floatfix,nolongbibliography]{revtex4-2}

\usepackage{graphicx}
\usepackage{dcolumn}
\usepackage{bm}
\usepackage{hyperref}
\usepackage{braket}
\usepackage{mathtools}
\usepackage{cleveref}

\begin{document}


\title{Low-energy model for doped graphene nanoribbons}

\author{J. Ferrer$^{1,2,3}$}
\author{A. García-Fuente$^{1,2}$}
\author{Y. Yang$^{3}$}
\author{S. Volosheniuk$^{3}$}
\author{H. S. J. van der Zant$^{3}$}

\affiliation{$^1$ Departamento de Física,  Universidad de Oviedo,  33007 Oviedo, Spain}
\affiliation{$^2$ Centro de Investigación en Nanomateriales y Nanotecnología, CSIC, 33940 El Entrego, Spain}
\affiliation{$^3$ Kavli Institute of Nanoscience, Delft University of Technology, Delft, The Netherlands}

\date{\today}


\begin{abstract}


We analyse in this article the many-body behavior of free-standing doped graphene nanoribbons 
where the chemical potential lies inside the bulk single-particle bands.
We perform an exact mapping from both an extended and an on-site Hubbard model of the ribbons
to a Kanamori model, which includes ferromagnetic exchange and pair-hopping interactions. 
We determine the resulting Coulomb matrix elements analytically,
and identify their scaling behavior as a function of ribbon width and length.
We propose a low-energy version of the Kanamori Hamiltonian to address the response of the
ribbons to external fields, with a view to their use as transport channels in nanoelectronics.
We find that the model and the proposed ribbon parameters can produce open-shell, high-spin 
many-body states that can lead to shell- and spin-blockade responses.
\end{abstract}

\maketitle


\section{Introduction}


Graphene was originally envisioned as the dreamed-of electronic material~\cite{Geim09}. The vision 
included the realization of nano-electronic devices such as transistors or quantum dots, frequently
based on graphene nanoribbons (GNRs) \cite{Westervelt09}. However, the expected revolution was hindered 
by a variety of factors, among which are the difficulties in producing atomically-precise GNRs~\cite{Cai10}.
Recently, atomically-precise GNRs have been grown in ultra-high vacuum chambers and transferred to substrates 
with metal electrodes. Transport through single- or few-nanoribbon devices has recently been measured, 
showing single-electron transistor and quantum dot behavior~\cite{Lawrence22,Zhang23a,Zhang23b,Niu23}.
In understanding their behavior, it is important to realize that graphene nanoribbons are different from 
carbon nanotubes because $\mathbf{\pm k}$ wave-vectors are
mixed by boundary conditions, so valleys are not good quantum numbers. Spin-splitting 
is expected to be absent in graphene nanoribbons except when the chemical potential is pinned around 
the edge-state energy window. Furthermore, both the intrinsic and Rashba spin orbit interactions are 
expected to be small like 
in bulk graphene, which exhibits a gap of about $25~\mu$eV~\cite{Laird15,Gmitra10,Zollner25}. 

In previous work \cite{Fuente23,Talkachov23}, we diagonalized analytically the tight-binding model of 
GNRs with arbitrary length and width, and showed that the existence or not, magnetic nature and energy of 
edge states 
in $N=5,\,7$ and 9 GNRs could be described by the use of an effective dimer Hubbard model.
The tight-binding solution was extended to the full Hubbard model of a GNR in 
Ref. \onlinecite{Ferrer25}. The Hubbard term was rewritten in the ribbon eigen-state basis, whereby a 
huge sum containing  operator products  $\hat{c}^\dagger_{1\sigma}\,
\hat{c}^\dagger_{2\sigma'}\,\hat{c}_{3\sigma'}\,\hat{c}_{4\sigma}$ was generated. The
sum was truncated by keeping only direct $1-4$, $2-3$ pairings, and the Coulomb integrals were calculated 
analytically. Explicit analytical mean-field equations for the paramagnetic, ferromagnetic and antiferromagnetic 
phases were deduced. The electronic band structure for narrow ribbons agreed with previous mean-field Hubbard 
model simulations \cite{Rossier08,Jung09}. The band structure for bulk-sized samples with edges did contain 
both the bulk Dirac cone as well as the edge states band at the same time, something that had not been achieved before.

In this article, we consider free-standing finite-length graphene nanoribbons (GNR). 
The ribbons are doped and possibly subject to a gate voltage so that the chemical potential 
lies inside the bands of bulk states. We provide an exact mapping
of the extended Hubbard model of GNRs \cite{Wehling11,Schuller13,Rossner15} to a Kanamori 
Hamiltonian \cite{Kanamori1963}, which is written in terms of the single-particle bulk 
eigen-states of the ribbon. 
We thus show that Hubbard interactions lead not only to direct Coulomb integrals among these single-particle bulk 
eigen-states, but also to ferromagnetic exchange and pair hopping terms. The pristine
extended Hubbard Hamiltonian is written in a real-space site basis; in contrast the Kanamori Hamiltonian
is written in the basis of single-particle eigen-states. This new Hamiltonian can be useful 
for the prediction or the analysis of  phenomena that address given energy windows. 
An example are recent transport experiments of single- or few-ribbon 
junctions ~\cite{Lawrence22,Zhang23a,Zhang23b,Niu23,Zhang26}, that have uncovered their quantum dot response. In these 
low-temperature experiments, a gate voltage tunes the chemical potential, and a narrow energy-window of 
bulk ribbon eigen-states is addressed. Detailed stability diagram featuring Coulomb diamonds and excited-state
lines can be drawn by sweeping both the bias and gate voltage in those experiments. 

We will discuss in this article how the new Hamiltonian allows us to address cleanly narrow 
energy-windows of eigen-states, whose position is given by the chemical potential and whose width 
is determined by and can be adapted to the number of single-particle states that are half-filled. 
The resulting low-energy Hamiltonian possesses a small enough Fock space that can be analized exactly 
and in detail either analytically or numerically. We find that the Kanamori exchange interactions 
are ferromagnetic and thus produce high-spin open-shell many-body states in doped GNR. Several theorems preclude 
the existence of saturated ferromagnetism for the Hubbard model \cite{Lieb62,Tasaki98}, but do predict 
high-spin (ferrimagnetic) ground states that fulfill specific conditions \cite{Pieri96,Tasaki98}. We 
find equivalent conditions for the high-spin ground states described below. Similarly, some
studies predict mesoscopic Stoner instabilities in open quantum dots \cite{Burmistrov20} as we do here.

The low-energy Hamiltonian approach is complementary to describing GNRs by Density Functional theory (DFT) 
simulations. This is because the Coulomb parameters introduced in the initial extended Hubbard model can and
must be deduced from DFT \cite{Wehling11,Schuller13,Rossner15,Fuente23}. On the other hand, correlated 
electron phenomena like Coulomb blockade or Kondo physics cannot be described correctly by DFT 
simulations. The reason is that DFT is a one-body theory so that even the exact functional does not 
describe the energy position and number of many-body quasi-particle poles  \cite{Carrascal12a,Carrascal15}.
A DFT-based analysis of Coulomb blockade phenomena delivers a wrong number and positioning of Coulomb 
diamonds, and of excited state lines.

\begin{figure}[ht]  \centering
  \includegraphics[width=0.95\columnwidth]{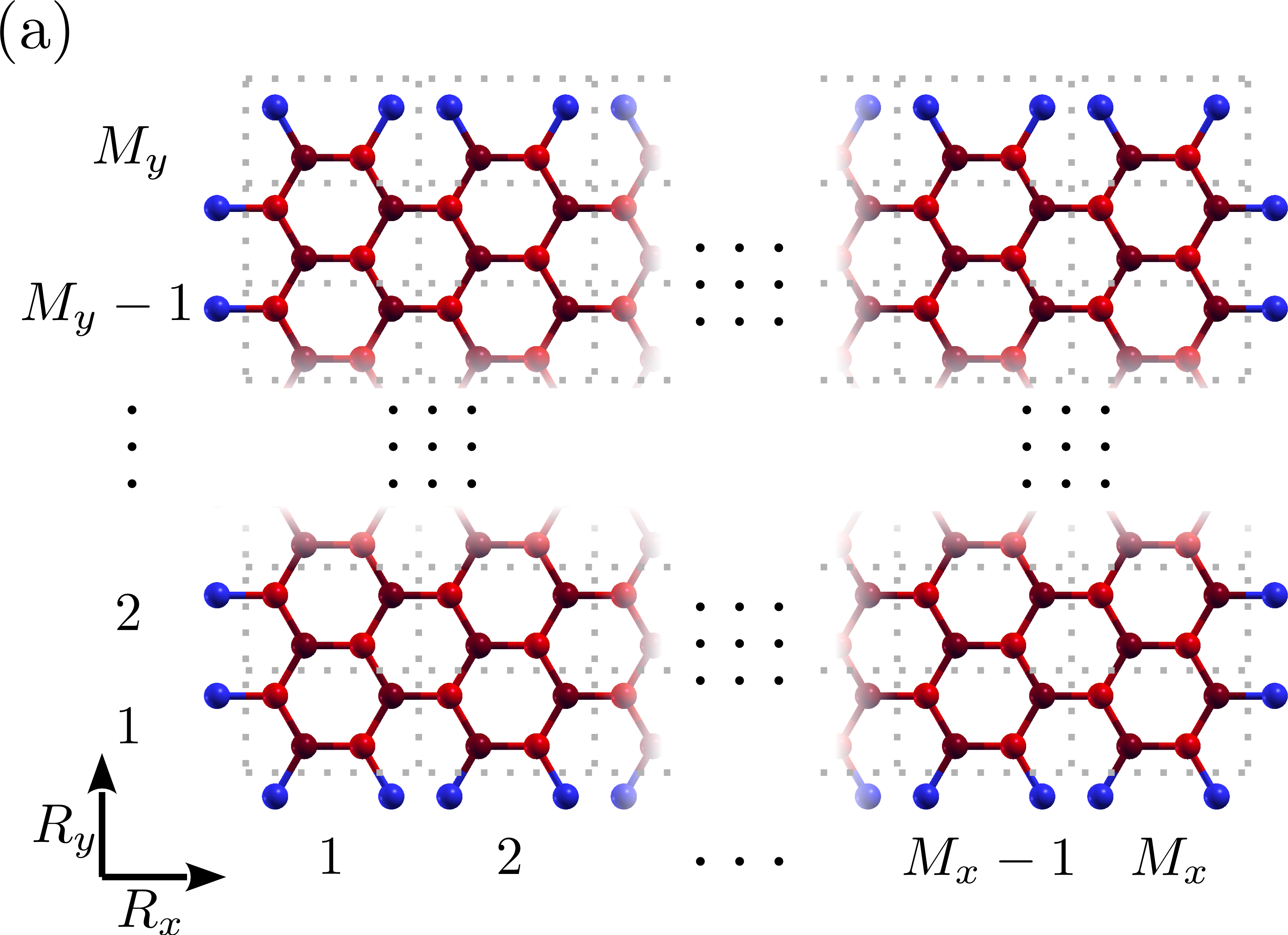}
  \caption{Real graphene atoms are indicated by red circles. 
  Fake atoms, whose wave-function coefficients are set to zero, are indicated by blue circles. 
  Each unit cell is surrounded with a gray dotted box. }
  \label{Figure:geometry_Ur}
\end{figure}

We will make this article self-contained and will therefore make use of some results that have been published
previously \cite{Fuente23,Ferrer25}. The layout of the article is as follows.
Section II summarizes the analytical solution of the tight-binding model of graphene ribbons of
arbitrary width and length \cite{Fuente23}. Section III explains the extended Hubbard model
of bulk graphene and introduces it for graphene nanoribbons. This section also shows how
the extended Hubbard Hamiltonian can be mapped to a Kanomori model \cite{Kanamori1963}, and
determines the Coulomb matrix elements. Section IV introduces an onsite Hubbard model and
fits the $U$ parameter using Density Functional Theory (DFT) simulations. The Coulomb 
matrix elements are vastly simplified here, and found to obey simple relationships among them,
and a remarkable size-scaling law. The section also discusses analytical mean field solutions
that were found earlier \cite{Ferrer25}. Section V uses the previous results to 
propose a low-energy Kanamori Hamiltonian to describe the open-shell and spin 
behavior of doped GNR. The low-energy Hamiltonian is solved for the case of two and three shells, 
and an specific example using a $N=9$ armchair ribbon is discussed. Section VI introduces 
a transfer Hamiltonian and discusses transition rates in the Sequential Electron Tunneling
regime. The section thus show how the model can lead to shell- and spin-blockade
in electrical transport experiments using the ribbons as channels connecting two leads.
A short conclusion closes the article.


\section{Tight-binding model of graphene ribbons}


We consider nanoribbons whose lengths and widths are $M_x$ 
and $M_y=\frac{N+1}{2}$, according to the convention set in Fig. 1. Unit cells are rectangles
having four atoms.
We measure lengths along the cartesian X- and Y-axes in units of $\sqrt{3}\,c$ and $c$, respectively, where $c=2.46$ \AA\ is graphene's lattice constant. Cell coordinates
are ${\bf R}=(n_x,\,n_y)$ with $n_{x/y}=1,\,2,\dots,\,M_{x/y}$.
Furthermore, we address for simplicity ribbons where no edge states 
contribute to the physics. This condition happens in both armchair and zigzag ribbons whenever 
the effective doping or the gate voltage are high enough that the chemical potential 
lies inside the conduction or valence bands. 

The tight-binding Hamiltonian is constructed by placing a Wannierized 
$p^z$ orbital
at each site $\bf{i}=\bf{R}+\bf{d}_{A/B}$ of the lattice with unit cell coordinates $\bf{R}$ 
and intra-cell coordinates $\bf{d}_{A/B}$. The corresponding states are represented by the 
kets $\ket{\bf{i}}=\ket{\bf{R}+\bf{d}_{A/B}}$. The second-quantized Hamiltonian is
\begin{equation}
    \hat{H}^0=-t\,\sum_{\braket{\bf{i},\bf{j}},\sigma}\,\hat{r}_{\bf{i}\sigma}^\dagger\,\hat{r}_{\bf{j}\sigma}   
\end{equation}
where the sum is restricted to nearest neighbor sites. The linear operator 
$\hat{r}^\dagger_{\bf{i}\sigma}$ creates an electron with spin $\sigma$ at the ribbon 
site $\bf{i}$, and $\hat{n}_{i\sigma}=\hat{r}_{i\sigma}^\dagger\,\hat{r}_{i\sigma}$ is the corresponding number 
operator, with $\hat{n}_i=\sum_\sigma\,\hat{n}_{i\sigma}$. 

The ribbon boundary conditions
define the set of allowed wave-vectors, that we define as 
${\bf \bar{k}}=(\bar{k}_x,\,\bar{k}_y)=(\sqrt{3}\,c\,k_x,\,c\,k_y)$.
Here, $\bar{k}_y=k_m=\pi\,\frac{m}{M_y}$ with $m=1,\dots,M_y$. The wave-vectors $\bar{k}_x=k_{m\alpha}$,
with $\alpha=1,\dots,\,2\,M_x$, must be computed numerically and are not spaced evenly. 
The subset $k_{m\,2 M_x}$ with $k_m \ge \frac{2}{3}\,\pi$
corresponds to edge ribbon states \cite{Fuente23,Ferrer25} and are discarded, while all the rest label bulk ribbon states and are kept.
The Hamiltonian $H^0$ can be diagonalized analytically for any non-chiral finite-length ribbon 
\cite{Fuente23}. The site operators can be decomposed in terms of bulk
eigen-state operators as follows
\begin{eqnarray}
\label{equation:decomposition}
\hat{r}_{{\bf i} \sigma}^\dagger&=& \sum_{m \alpha \tau} \,
\hat{r}_{m \alpha \tau \sigma}^\dagger\braket{\phi_{m \alpha \tau}\,|\,{\bf i}}
\\
\braket{\bf{i}\,|\,\phi_{m\alpha\tau}}&=&\sum_{R_{x,y}=1}^{M_{x,y}}\,
\frac{2\,\sin{k_m\,(R_y-d_i)}}{({\cal M}_x\,M_y\,A_{m\alpha})^{1/2}}\,\times\\
&&\times\begin{pmatrix}-\tau\,(-1)^\alpha\,\sin(k_{m\alpha}\,(R_x-d_i))
\\\sin(k_{m\alpha}\,(M_x+1-(R_x+d_i)))
\end{pmatrix}\nonumber
\end{eqnarray}
where ${\cal M}_x=4\,M_x+1$, and $\tau=\pm 1$ is a conduction/valence band index, and $d_1=0$ and $d_2=1/2$. $\{m,\,\alpha,\,\tau,\,\sigma\}$ is a complete set of quantum numbers in the ribbon. The 
eigenstates are spin-degenerate, and we call the pair $(m\,\alpha\,\tau\uparrow),\,(m\,\alpha\,\tau\downarrow)$ a {\it shell} from now on. The 
normalization factor is
\begin{eqnarray}
  A_{m \alpha}&=&F^1(k_{m\alpha})-\frac{\delta_{k_m,\pi}}{{\cal M}_x}\,F^2(k_{m\alpha})\\
  F^1(k)&=&1-\frac{\sin{({\cal M}_x\,k/2)}}{{\cal M}_x\,\sin({k/2})}\\
  F^2(k)&=&1-\frac{\cos{({\cal M}_x\,k/2)}}{{\cal M}_x\,\cos{(k/2)}}
\end{eqnarray}
The diagonalized Hamiltonian and the single-particle eigen-energies are 
\begin{eqnarray}
 \hat{H}^0&=&\sum_{m\,\alpha\,\tau}\,\epsilon_{m\alpha\tau}^0\,\hat{n}_{m\alpha\tau}\\
\epsilon_{m\alpha\tau}^0&=&\tau\,\sqrt{1+\Delta_m^2+2\Delta_m\cos{\frac{k_{m\alpha}}{2}}}
\end{eqnarray}
where $\Delta_m=2\,\cos{\frac{k_m}{2}}$. 


\section{Extended Hubbard model of graphene}


\subsection{Bulk graphene}
The extended Hubbard model is constructed by adding to $ \hat{H}^0$ the following interaction Hamiltonian: 
\begin{eqnarray}
    \label{equation:extended}
    \hat{H}^I&=&U^0\,\sum_{\bf{i}}\,\hat{n}_{\bf{i}\uparrow}\,\hat{n}_{\bf{i}\downarrow}+
                 \sum_{n,\bf{i},\bf{\delta}^n}\,U^n\,\hat{n}_{\bf i}\,\hat{n}_{\bf{i} +\bf{\delta}^n}
\end{eqnarray}
The second sum extends to the different atomic neighbor shells $n$, while $\delta^n$ runs 
over all the atoms in the $n^{th}$-neighbor shell of atom $i$. 
The on-site and extended Hubbard 
parameters $U^0$ and $U^n$ correspond to the effective Coulomb interactions among electrons residing in the wannierized 
$p^z$ orbitals. These interactions correspond to the bare Coulomb interaction among $p^z$ orbitals, screened by all 
the remaining higher energy orbitals, and eventually by the presence of a substrate and
perhaps an overlayer. $U^0$ 
and $U^n$ have been obtained for free-standing graphene and for bilayer graphene encapsulated or deposited on idealized 
metallic substrates in references \onlinecite{Wehling11,Rossner15}. The reported Hubbard parameters $U^n$ for free-standing 
graphene can be rewritten in terms of the distance $r$ between two sites and are very well fitted to 
the law \cite{Wehling11,Rossner15}
\begin{equation}
\label{equation:bulklaw}
    U(r)=\frac{10.3}{1+0.6\,r}\,\mathrm{eV}
\end{equation}
where $r$ must be given in \AA. In addition, reference \onlinecite{Schuller13} has analyzed whether an on-site Hubbard model can describe free-standing 
graphene. The article finds that this is the case and reports an effective on-site $U^0\approx 4.3$ eV, except for extreme 
doping cases very far away from half-filling.


\subsection{Graphene ribbons}
We rewrite in this section the interaction Hamiltonian $\hat{H}^I$ in the basis of single-particle eigen-states of the ribbon. 
Inserting the decomposition \ref{equation:decomposition} into Eq. \ref{equation:extended}, we find a large sum of terms of the  type
\begin{equation}
\hat{r}_{m_1 \alpha_1 \tau_1 \sigma_1}^\dagger\,\hat{r}_{m_2 \alpha_2 \tau_2 \sigma_2}^\dagger\,
\hat{r}_{m_3 \alpha_3 \tau_3 \sigma_3}\,\hat{r}_{m_4 \alpha_4 \tau_4 \sigma_4}
\end{equation}
that are multiplied by Coulomb integrals. However, many of these integrals are zero because
momentum conservation implies the condition $k_{m_1\alpha_1}+k_{m_2\alpha_2}=k_{m_3\alpha_3}+k_{m_4\alpha_4}$. 
The reason for the cancellation is that the wave-vectors $k_{m\alpha}$ are not separated evenly for a ribbon, 
so the condition is not fulfilled except when $\mathbf{k}_1=\mathbf{k}_3$ and 
$\mathbf{k}_2=\mathbf{k}_4$, or when $\mathbf{k}_1=\mathbf{k}_4$ and $\mathbf{k}_2=\mathbf{k}_3$.
In addition, the two single-particle bulk states having the same $k_{m\alpha}$ and opposite $\tau$ are effectively 
unentangled because they possess very different energies. 
As a consequence, momentum conservation allows us to rewrite 
the following effective Hamiltonian for bulk ribbons states, which is of the Kanamori type \cite{Kanamori1963}:
\begin{widetext}
\begin{equation}
\label{equation:kanamori}
\hat{H}=\sum_{a}\,\epsilon^0_{a}\,\hat{n}_{a}+
\sum_{a}\,V_{a}\,\hat{n}_{a\uparrow}\,\hat{n}_{a\downarrow}
+\sum_{a\neq b}\,\left(V_{a b}^n\,\hat{n}_a\,\hat{n}_{b}
-V_{a b}^S\,\mathbf{\hat{S}}_{a}\cdot\mathbf{\hat{S}}_{b}+
V_{a b}^\Delta\,\frac{\hat{\Delta}_a^\dagger\,\hat{\Delta}_{b}+
\hat{\Delta}_{b}^\dagger\,\hat{\Delta}_{a})}{2}\right)
\end{equation}
\end{widetext}
where $a$ and $b$ are shorthands for the single-particle quantum numbers $(m\,\alpha\,\tau)$ and $(m'\,\alpha'\,\tau')$. 
$\mathbf{\hat{S}}_a=\hat{r}_{a\sigma}^\dagger\frac{\mathbf{\tau}_{\sigma\sigma'}}{2}\hat{r}_{a\sigma'}$ are spin operators while
$\hat{\Delta}_a^\dagger=\hat{r}_{a\uparrow}^\dagger\,\hat{r}_{a\downarrow}^\dagger$
are pair-hopping operators. The Hamiltonian parameters possess the following explicit expressions
\begin{eqnarray}
\label{Equation:matrix_elements_1}
 \epsilon_a&=&\tau \epsilon^0_a + V_{aa}\\
 V_a&=&V_{a a}^0+V_{aa}\\
 V_{a b}^n&=&\frac{V_{a b}^0}{4}+V_{a b}-\frac{J_{a b}}{2}\\
 V_{a b}^S&=&V_{a b}^\Delta=V_{a b}^0+2\,J_{a b}\ V_{a b}^\Delta
 \end{eqnarray}
 The direct and exchange matrix elements can be written as follows:
 \begin{eqnarray}
 V_{ab} &=&V_{ab}^1+V_{ab}^2+V_{ab}^3+\ldots\\
   \label{Equation:matrix_elements_2}
 J_{ab} &=&J_{ab}^1+J_{ab}^2+J_{ab}^3+\ldots
\end{eqnarray}
where each $V_{ab}^n$ and $J_{ab}^n$ derives from the Hubbard terms $U^n\,\hat{n}_{\bf i}\,\hat{n}_{\bf{i} +\bf{\delta}^n}$ that 
connect sites separated by $n-$shells. The onsite-derived interactions are
\begin{eqnarray}
\label{equation:onsite}
  V_{m\alpha;m'\alpha'}^0&=&2\,C\,U^0\,\sum_{i=1,\,2}
  \sum_{n_x}\,\sin^2(k_{m\alpha}\,n_x^i)\,\sin^2(k_{m'\alpha'}\,n_x^i)\,\times\nonumber\\
  &&\times\sum_{n_y}\,\sin^2(k_{m}\,n_y^i)\,\sin^2(k_{m'}\,n_y^i)\\
  C&=&\frac{16}{{\cal M}_x^2\,M_y^2\,A_{m\alpha}\,A_{m'\alpha'}}
\end{eqnarray}
with vectors $\mathbf{n}^1=(n_x,\,n_y)$ and $\mathbf{n}^2=(n_x-1/2,\,n_y-1/2)$.
The explicit expressions for the shell-dependent matrix elements depend on whether 
$U^n$ connects same- or opposite-sublattices. 
When $U^n$ connects same-sublattice shells with neighbor vectors ${\boldsymbol \delta}^n$, we 
find  
\begin{widetext}
\begin{eqnarray}
V_{m\alpha;m'\alpha'}^n&=&C\,U^n\,\sum_{i=1,\,2}\,\sum_{\bf{\delta^n}}
  \sum_{n_x}\,\sin^2(k_{m\alpha}\,n_x^i)\,\sin^2(k_{m'\alpha'}(n_x^i+\delta_x^n))\,
  \,\sum_{n_y}\,\sin^2(k_{m}\,n_y^i)\,\sin^2(k_{m'}(n_y^i+\delta_y^n))\\
J_{m\alpha;m'\alpha'}^n&=&C\,U^n\,\sum_{i=1,\,2}\,\sum_{\bf{\delta^n}}
  \sum_{n_x}\,\sin(k_{m\alpha}\,n_x^i)\,\sin(k_{m'\alpha'}\,n_x^i)
  \sin(k_{m\alpha}(n_x^i+\delta_x^n))\, \sin(k_{m'\alpha'}(n_x^i+\delta_x^n))
  \times\nonumber\\
  &&\times\,\sum_{n_y}\,\sin(k_{m}\,n_y^i)\,\sin(k_{m'}\,n_y^i)\,
  \sin(k_{m}(n_y^i+\delta_y^n))\sin(k_{m'}(n_y^i+\delta_y^n))
\end{eqnarray}
In contrast, when $U^n$ connects opposite-sublattice shells with neighbor vectors 
${\boldsymbol \delta}^n$, we find  
\begin{eqnarray}
    V_{m\alpha;m'\alpha'}^n&=&C\,U^n\,\sum_{i=1,\,2}\,\sum_{\bf{\delta^n}}
  \sum_{n_x}\,\sin^2(k_{m\alpha}\,n_x^i)\,\sin^2(k_{m'\alpha'}(M_x+1-(n_x^i+\delta_x^n)))\,
  \,\sum_{n_y}\,\sin^2(k_{m}\,n_y^i)\,\sin^2(k_{m'}(n_y^i+\delta_y^n))\\
J_{m\alpha;m'\alpha'}^n&=&C\,U^n\,\sum_{i=1,\,2}\,\sum_{\bf{\delta^n}}
  \sum_{n_x}\,\sin(k_{m\alpha}\,n_x^i)\,\sin(k_{m'\alpha'}\,n_x^i)
  \sin(k_{m\alpha}(M_x+1-(n_x^i+\delta_x^n)))\, \sin(k_{m'\alpha'}(M_x+1-(n_x^i+\delta_x^n)))
  \times\nonumber\\
  &&\times\,\sum_{n_y}\,\sin(k_{m}\,(n_y^i))\,\sin(k_{m'}\,(n_y^i))\,
  \sin(k_{m}(n_y^i+\delta_y^1))\sin(k_{m'}(n_y^i+\delta_y^1))
\end{eqnarray}
\end{widetext}
Finally, the Hubbard parameters $U^n$ are well-fitted by Eq. \ref{equation:bulklaw} for free-standing bulk 
graphene \cite{Wehling11,Rossner15}, but are unknown for graphene ribbons. A possible
strategy consists of borrowing and using the bulk parameters in Eq. \ref{equation:bulklaw}.


\section{On-site Hubbard model of graphene ribbons}


\subsection{Determination of the onsite Hubbard parameter $U^0$}
We follow in the rest on this article a second strategy, which consists of reducing the
parameters of the extended Hubbard model down to an onsite renormalized interaction $U^0$.
This strategy was implemented for bulk graphene in Ref. \onlinecite{Schuller13}, where the
authors found a renormalized $U^0\simeq 4.3$ eV. We attempt a similar task in this section
for graphene ribbons. We make use of the fact that DFT is a mean field theory 
akin in spirit to performing a Hartree decomposition of the on-site Hubbard model. 
We perform DFT and Hartree decompositions of the onsite Hubbard model for armchair ribbons with
$M_y=4,\,5,\,9$ (corresponding to $N=7,\,9,\,17$) and lengths ranging from $M_x=2$ up to 
infinity. We find that an onsite Hubbard $U^0=4.9$ eV fits the DFT bulk gap and the 
lowest-lying electron and hole excitations (also called HOMO-1, HOMO, LUMO and LUMO+1)
rather closely, as demonstrated in Tables I, II and III. We therefore choose $U^0=4.9$ eV
for the remaining of this article.

\setlength{\tabcolsep}{3pt}
\begin{table}[ht]
    \centering
    \begin{tabular}{|c|c|c|c|c|c|}
    \hline
         $M_x$ & DFT H/L & MFH H/L & DFT H-1/L+1 & MFH H-1/L+1\\
   \hline
    2    &  0.943 & 0.958 & 3.809 & 3.775 \\
    \hline
    3    &  0.596 & 0.600 & 3.127 & 3.199 \\
    \hline
    4    &  0.556 & 0.534 & 2.545 & 2.650 \\
    \hline
   5    &  0.546  & 0.504 & 2.210 & 2.237 \\
   \hline
   6    &  0.545  & 0.491 & 2.017 & 1.975 \\
   \hline
   7    &  0.546  & 0.484 & 1.898 & 1.801 \\
   \hline
   8    &  0.547  & 0.482 & 1.817 & 1.681 \\
   \hline
   9    &  0.547  & 0.481 & 1.760 & 1.595 \\
   \hline
   10   &  0.547  & 0.480 & 1.719 & 1.532 \\
   \hline
    \end{tabular}
    \caption{HOMO-LUMO gaps in eV for $M_y=4$, $N=7$ armchair GNRs of different lengths $M_x$, as obtained from our Density 
    Functional Theory (DFT) and antiferromagnetic mean-field Hubbard (MFH) calculations with $U^0=4.9$ eV. The 
    infinite DFT gap is 1.559 eV.}
\end{table}

\setlength{\tabcolsep}{3pt}
\begin{table}[ht]
    \centering
    \begin{tabular}{|c|c|c|c|c|c|}
    \hline
         $M_x$ & DFT H/L & MFH H/L & DFT H-1/L+1 & MFH H-1/L+1\\
   \hline
    2    &  0.699 & 0.745 & 2.760 & 2.662 \\
    \hline
    3    &  0.663 & 0.678 & 2.075 & 2.063 \\
    \hline
    4    &  0.660 & 0.672 & 1.713 & 1.749 \\
    \hline
   5    &   0.659 & 0.658 & 1.498 & 1.558 \\
   \hline
   6    &   0.658 & 0.658 & 1.357 & 1.431 \\
   \hline
   7    &   0.657 & 0.657 & 1.258 & 1.341 \\
   \hline
   8    &   0.658 & 0.657 & 1.186 & 1.275 \\
   \hline
   9    &   0.657 & 0.657 & 1.133 & 1.225 \\
   \hline
   10    &  0.658 & 0.657 & 1.088 & 1.186 \\
   \hline
    \end{tabular}
    \caption{Same as in Table I, with $M_y=5$, $N=9$. The 
    infinite DFT gap is in this case 0.830 eV.}
\end{table}

\setlength{\tabcolsep}{3pt}
\begin{table}[ht]
    \centering
    \begin{tabular}{|c|c|c|c|c|c|}
    \hline
         $M_x$ & DFT H/L & MFH H/L & DFT H-1/L+1 & MFH H-1/L+1\\
   \hline
    2    &   0.793    & 0.706 & 0.848 & 0.931  \\
    \hline
    3    &    0.624   &  0.609 & 0.884 & 0.919 \\
    \hline
    4    &    0.592   & 0.579 & 0.892 & 0.914 \\
    \hline
   5    &   0.585    & 0.570 & 0.804 & 0.807 \\
   \hline
   6    &   0.584    & 0.567 & 0.666 & 0.678 \\
   \hline
   7    &   0.566    & 0.566 & 0.583 & 0.585 \\
   \hline
   8    &   0.490    & 0.514 & 0.584 & 0.566 \\
   \hline
   9    &   0.431    & 0.458 & 0.584 & 0.566 \\
   \hline
   10    &   0.384    & 0.414 & 0.584 & 0.566 \\
   \hline
    \end{tabular}
    \caption{Same as in Table I, with $M_y=9$, $N=17$. The 
    infinite DFT gap is 0.114 eV.}
\end{table}

Notice that the matrix elements in Equs. \ref{Equation:matrix_elements_1}-\ref{Equation:matrix_elements_2} simplify for the 
on-site Hubbard model and fulfill $4\,V_{ab}^n=V_{ab}^S=2\,V_{ab}^\Delta=V^0_{ab}$. 
so that the
Hamiltonian in Eq. \ref{equation:kanamori} simplifies to
\begin{eqnarray}
\hat{H}&=&\sum_{a}\,\left(\epsilon^0_{a}\,\hat{n}_{a}+
V_{aa}^0\,\hat{n}_{a\uparrow}\,\hat{n}_{a\downarrow}\right)+\\
&&+\sum_{a\neq b}\,V_{a b}^0\,\left(\frac{\hat{n}_a\,\hat{n}_{b}}{4}
-\mathbf{\hat{S}}_{a}\cdot\mathbf{\hat{S}}_{b}+\frac{
\hat{\Delta}_a^\dagger\,\hat{\Delta}_{b}+
\hat{\Delta}_{b}^\dagger\,\hat{\Delta}_{a}}{2}\right)\nonumber
\end{eqnarray}
A further advantage of this model is that the on-site matrix elements in Eq. \ref{equation:onsite} 
can be computed analytically \cite{Fuente23,Ferrer25}. This allows us to determine them even for ribbons 
with sizes of the order of micrometers. The expressions are
\begin{eqnarray}
V^0_{m\alpha,m'\alpha'}&=&{\cal D}
\,\,\frac{\delta^+_{m m'}\,C^{B,+}_{m\alpha,m'\alpha'}+\delta^-_{m m'}\,C^{B,-}_{m\alpha,m'\alpha'}/
{\cal M}_x}{\Lambda^\phi_{m\alpha}\,\Lambda^\phi_{m'\alpha'}}\nonumber\\
\end{eqnarray}
where
\begin{eqnarray}
{\cal D}&=&\frac{U^0}{2\,{\cal M}_x\,M_y}\\
\delta_{m m'}^+&=&1+\frac{1}{2}\left(\delta_{m,m'}+\delta_{m,M_y} \delta_{m',M_y}\right)\\
\delta_{m m'}^-&=&\frac{1}{2}\,\delta_{m+m',M_y}-\left(\delta_{m,M_y}+\delta_{m',M_y}\right)\\
C^{B,+}_{m\alpha,m'\alpha'}&=&2\,F^1(k_{m \alpha})+2\,F^1(k_{m' \alpha'})-F^1(k_+)-F^1(k_{-})\nonumber \\\\
C^{B,-}_{m\alpha,m'\alpha'}&=&2\,F^2(k_{m \alpha})+2\,F^2(k_{m' \alpha'})-F^2(k_{+})-F^2(k_{-})\nonumber\\
\end{eqnarray}
with $k_{\pm}=k_{m\alpha}\pm k_{m'\alpha'}$.

\begin{figure}[ht]  \centering
  \includegraphics[width=0.95\columnwidth]{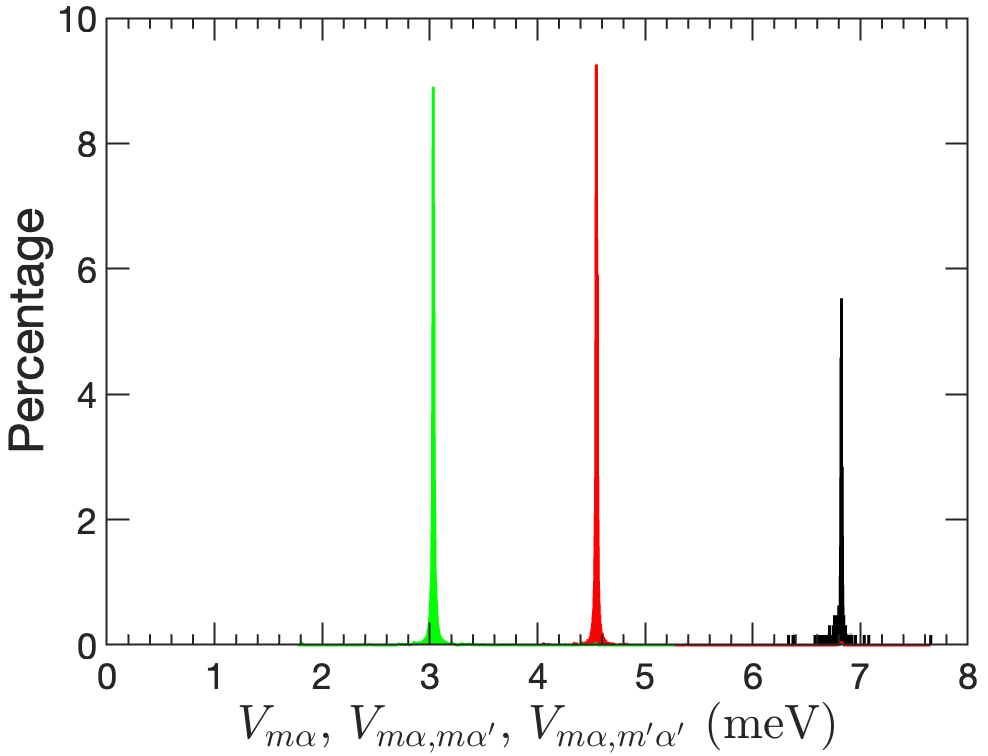}
  \caption{Histograms of the Coulomb matrix elements of an armchair ribbon of width $M_y=5$ 
  ($N=9$) and  length $M_x=80$ ($L = 34.1$ nm), as a function of energy measured in meV, for 
  the four bands $m=1,...,4$, and all wave-vectors $\alpha=1,...,\,2\,M_x-1$.
  $V_{ma}$ is shown in black color; the peak is placed at 6.8 meV.
  Intra-band $V_{m\alpha,m\alpha'}$  and inter-band $V_{m\alpha,m'\neq m\alpha'}$ are shown
  in red and green colors, respectively. The two peaks are placed at 4.53 and 3.02 meV.}
  \label{Figure:histograms}
\end{figure}

Upon computing numerically the matrix elements $V^0_{m\alpha;m'\alpha}$ for ribbons of different sizes, we come 
to the conclusion that they collapse into three types that only depend on the ribbon size: same-state 
integrals $V_{m\alpha}^0=V^0_{aa}={\cal U}(M_x,M_y)$, same-band integrals $V_{m\alpha,m\alpha'}^0={\cal V}(M_x,M_y)$ and 
the rest $V_{m\alpha,m'\alpha'}^0={\cal W}(M_x,M_y)$. Consequently, only three Coulomb integrals 
are needed for each ribbon size. An example of this behaviour for an armchair ribbon of size 
$M_y=5$ ($N=9$) and $M_x=80$ is shown in Fig. \ref{Figure:histograms}.
We have also found that these three Coulomb integrals follow closely the ratios 
${\cal V}=2/3\,{\cal U}$ and ${\cal W}=2/3\,{\cal V}$, up to an accuracy of less than 1 per cent,
as can also be inferred from the same figure. As a consequence, it suffices to determine 
one of the three, and we choose ${\cal U}$. We have fitted ${\cal U}$ for armchair ribbons 
with widths and lengths ranging from $M_y=3$ to 9 and $M_x=3$ to 110 respectively, and have 
found that the size-scaling relationship
\begin{eqnarray}
{\cal U} =\frac{3.19}{(1+1.156\,M_x)\, M_y} \,\mathrm{eV}.
\end{eqnarray}
is accurate up to one or two percent.

\begin{figure}[ht]  \centering
  \includegraphics[width=0.5\textwidth]{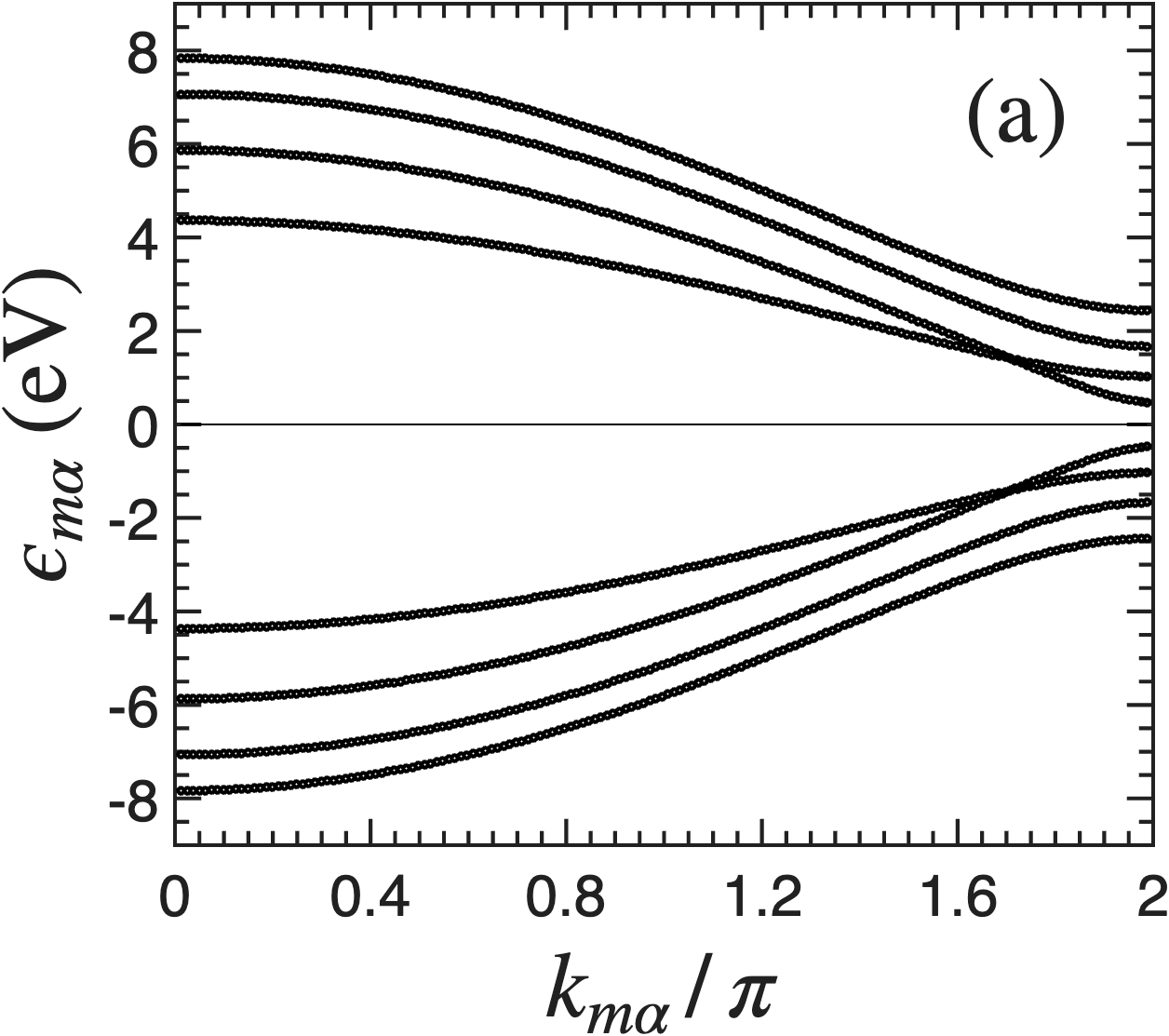}
  \caption{
  Bulk eigen-energies $\epsilon_{m\alpha}^{MF}$ of the antiferromagnetic mean field solution of an 
  armchair ribbon of width $M_y=5$  ($N=9$) and  length $M_x=80$ ($L = 34.1$ nm) \cite{Ferrer25}.
  These eigen-energies are discrete; they resemble bands because of the close spacing of the wave-vectors
  $k_{m\alpha}$ for this ribbon.}
  \label{Figure:bands}
\end{figure}


\subsection{Mean-field solution for undoped armchair ribbons}
\begin{figure}[ht]  \centering
  \includegraphics[width=0.22\textwidth]{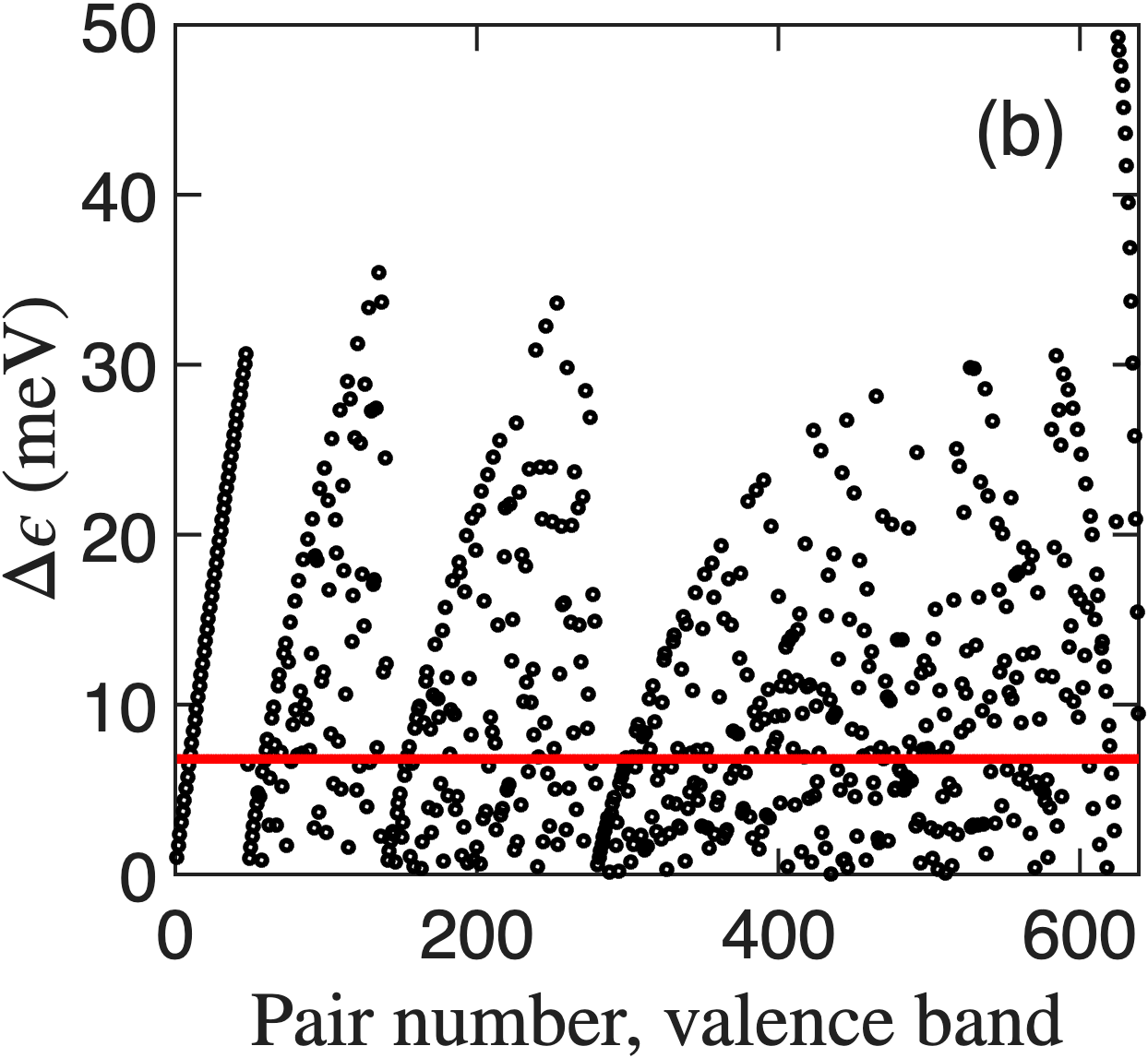}
  \includegraphics[width=0.24\textwidth]{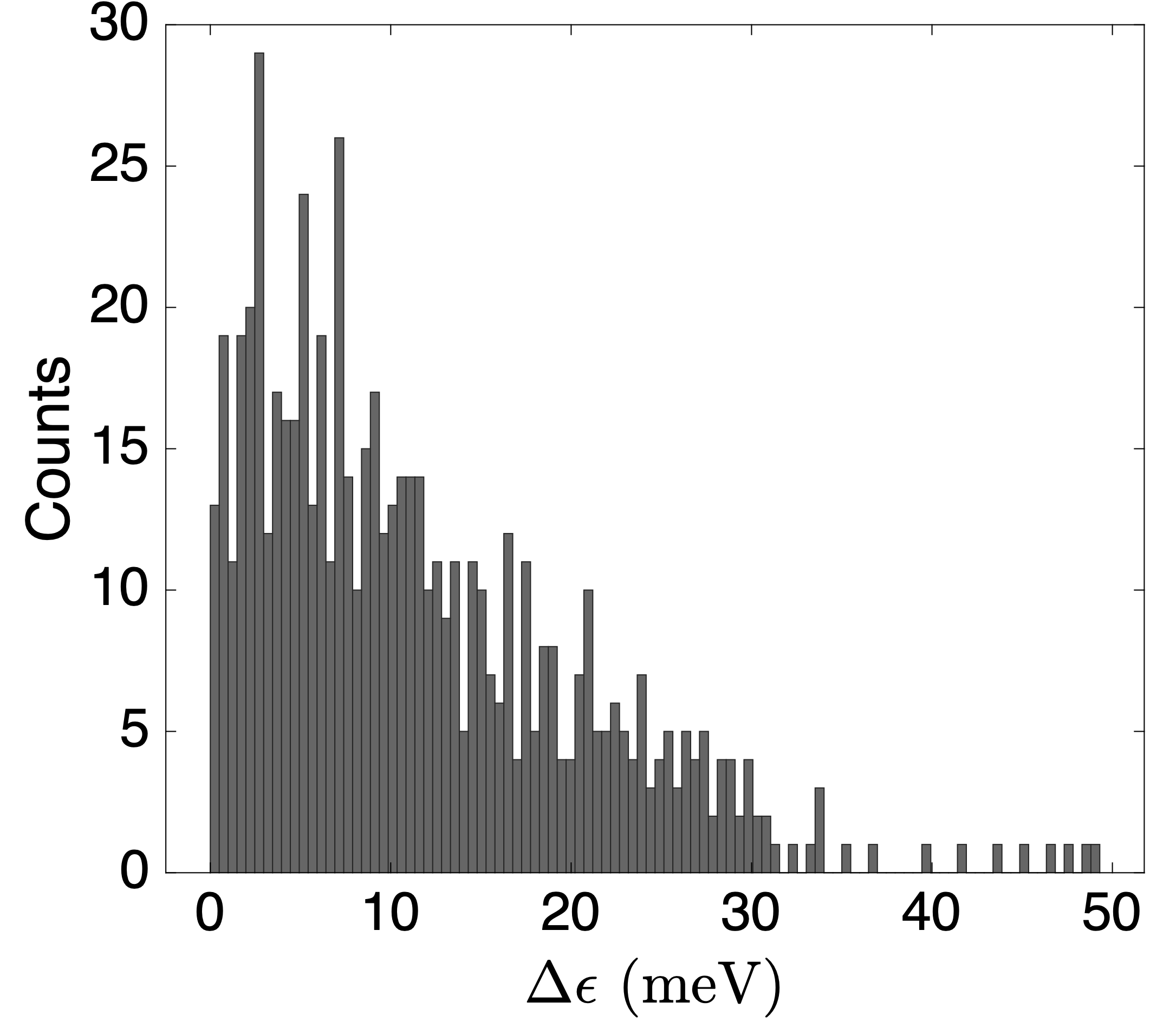}
  \caption{(a) Mean-field single-particle energy differences 
  $\Delta\epsilon_{ab}=\epsilon_{a}^{MF}-\epsilon_{b}^{MF}$ 
  between consecutive-energy 
  bulk states belonging to the valence band, all extracted from Fig. \ref{Figure:bands}. 
  The horizontal red line is placed at an energy equal to $V=6.8$ meV, which corresponds to the 
  Coulomb matrix element ${\cal U}$ of this ribbon. (b) Histogram of the same energy
  differences.}
  \label{Figure:addition_energies}
\end{figure}

We have performed the Kanamori mapping for undoped ribbons in a previous article \cite{Ferrer25}.
There, we kept both edge and bulk states but dropped the exchange and pair-hopping terms. 
We performed an analytical mean field treatment of the Hamiltonian and found paramagnetic, 
ferromagnetic and antiferromagnetic solutions. These solutions agreed with previous numerical calculations
carried out for infinite length ribbons, both for the bulk and for the edge bands. We also found that
the three solutions delivered the same properties for 
bulk states, and only differed on the nature and energies of the edge states. 
Specifically, mean-field bulk states kept a one-to-one relationship with their non-interacting
counterparts: they retained the quantum numbers $(m\,\alpha\,\tau\,\sigma)$ and were 
spin-degenerate. All in all, bulk states could still be classified by {\it shells} labeled by 
$(m\,\alpha\,\tau)$.  
We also found that the filled bulk mean field states undergo an energy shift 
\begin{eqnarray}
    \epsilon_{m\alpha}^0\rightarrow\epsilon_{m\alpha\sigma}^{MF}=\epsilon_{m\alpha}^0+
    \,{\cal U}\,n_{m\alpha-\sigma}+\sum_{\mathrm occ}
    V_{m\alpha;m'\alpha'}^0\,n_{m'\alpha'}\nonumber\\
\end{eqnarray}
where the sum runs over all occupied states. The energy shift is essentially the same for all
bulk states, and so it amounts to a rigid shift of all the bulk bands. These mean field bulk 
energies are illustrated in Fig. \ref{Figure:bands} for a $M_y=5$ ribbon of length $M_x=80$.
The mean-field eigen-energies can be used to determine the single-particle addition energies 
$\Delta \epsilon_{m\alpha,m'\alpha'}=\epsilon_{m\alpha}^{MF}-\epsilon_{m'\alpha'}^{MF}$ between 
consecutive-energy states.
Notice that the two single-particle energies can belong to the same band or two different bands.
The first case leads to a uniform pattern, while the second follows a random distribution. This 
behavior is illustrated in Fig. \ref{Figure:addition_energies} for the same ribbon, with
$M_y=5$ and $M_x=80$. Panel (a) shows $\Delta\epsilon_{ab}$, where the uniform patterns belonging
to same-band pairs can be appreciated; super-imposed to them, the random cloud of addition 
energies correspond to different-band pairs. Panel (b) shows the histogram of addition
energies. This histogram resembles a Wigner-Dyson random distribution, although we do not
pursue this issue further due to the limited number of counts. Instead, we would like
to highlight the large number of counts with $\Delta\epsilon_{ab}<{\cal U}\approx 6.8$
meV. This means that the electronic correlations of ribbons of these widths and lengths belong
to the intermediate regime, and may lead to interesting behavior beyond that expected for weak
electron correlations. This intermediate regime behavior will be exploited and analysed in 
the rest of this article.

\section{Low-energy Hamiltonian}


\subsection{Hamiltonian}
We remind that we analyse in this article situations where the chemical potential $\mu$ lies 
well within the bands of bulk single-particle states. 
With these conditions, bulk shells with energies sufficiently below $\mu$ are filled 
and inert, and we call them {\it closed shells}. Similarly,
bulk shells with energies sufficiently above $\mu$ are empty and inert, and similar 
words can be said for the ribbon edge shells, provided that they exist.
Inert means here that virtual transitions towards/from those states, triggered by either quantum or thermal 
fluctuations, have negligible probabilities. As a consequence, we expect that only a few shells will be active
for a given value of the chemical potential, and we call them  {\it open shells}. Therefore, when a gate voltage 
is swept, shells are opened and closed sequentially. 

We expect that a mean-field treatment should be sufficient to describe closed shells. 
However, shells that are close enough to the chemical potential might be open
or closed depending of the delicate interplay between kinetic and Coulomb energies.
We thus discard from the Kanamori Hamiltonian those closed shells whose mean-field energies are 
sufficiently far from the chemical potential and keep only a small number $O$ of shells with 
mean-field energies around $\mu$. We thus arrive to the following low-energy many-body
hamiltonian that addresses cleanly these $O$ putative open shells 
\begin{eqnarray}
  \label{Equation:H_0} 
    \hat{H}&=&\sum_{a\in O}\,\epsilon_a^{MF} \hat{n}_a+{\cal U}\,\sum_{a\in O}\,\,\hat{n}_{a\uparrow}\,\hat{n}_{a\downarrow}+\\
&&+\sum_{a,b\in O}\,V_{a b}^0\,\left(\,\frac{\hat{n}_a\,\hat{n}_{b}}{4}
-\mathbf{\hat{S}}_{a}\cdot\mathbf{\hat{S}}_{b}+
\frac{\hat{\Delta}_a^\dagger\,\hat{\Delta}_{b}+
\hat{\Delta}_{b}^\dagger\,\hat{\Delta}_{a}}{2}\right)\nonumber\\
\epsilon_a^{MF}&=&\epsilon_a^{0}+\sum_{b\in\mathrm occ}  V_{a b}^0\,n_{b}
\end{eqnarray}
where the same-state, opposite-spin interaction of the open shells is kept out of the mean-field treatment. The
Coulomb integrals $V_{ab}^0$ are equal to ${\cal V} =2/3 \,{\cal U}$ or ${\cal W}=(2/3)^2 \,{\cal U}$ depending on whether
$a$ and $b$ belong or not to the same band, respectively.

The Fock space is spanned in the single-particle number basis 
$\{|n_{i\sigma})=|n_{1\uparrow},\,n_{1\downarrow},\,...,\,n_{O\uparrow},\,n_{O\downarrow})\}$ and has dimension $4^O$. 
The many-body Hamiltonian commutes with the total number operator $\hat{N}$ and with the total spin operators
$\hat{S}_T^2$, $\hat{S}_T^z$.
As a consequence the Fock space can be classified into $2\,O+1$ sub-spaces labeled by $N=0,\dots,2\,O$. 
Each of these subspaces can be further split into subspaces with given $S$. We find that there exists
frequently more than one subspace with quantum numbers $(N,\, S)$. Therefore, an additional quantum number $R$ needs
to be added to identify unambiguously each subspace $(N,\,S,\,R)$. Each of these  subspaces contains $2\,S+1$
eigenstates. The highest spin subspace available occurs at half-filling $N=O$ and the spin equals $S_T=O/2$. 
The ribbon many-body eigen-states are decomposed in the single-particle basis as 
\begin{equation}
\label{Equation:rotation}
\ket{N\,\lambda}=\sum 
|n_{i\sigma})\,(n_{i\sigma} \ket{N\lambda}=\sum |n_{i\sigma})\,{\cal R}^{N}_{{n_{i\sigma}},\lambda}
\end{equation}
where $\lambda$ is a shorthand for the many body quantum numbers $(R,\,S,\,S^z)$, and where
the sum is restricted to the combinations that sum up to $N$. The expansion coefficients define
the matrices ${\cal R^N}$ that diagonalize the Hamiltonian within each $N$ subspace. 
The diagonalized Hamiltonian is
\begin{eqnarray}
\Hat{H}=\sum_{N,\lambda}\,E_{N,\lambda}\, \hat{\Gamma}_{N,\lambda}^\dagger\,  \hat{\Gamma}_{N,\lambda}
\end{eqnarray}
where the linear operator $\Gamma_{N\lambda}^\dagger$ creates the ribbon eigen-state $(N,\,\lambda)$. 

The low-energy many-body Hamiltonian describes adequately the competition between open- and
closed-shell behavior, beyond a mean field treatment. This competition is driven by the 
interplay between $\Delta\epsilon_{ab}=\epsilon_b^{MF}-\epsilon_a^{MF}$, 
${\cal U}$ and $V_{ab}^0$. We discuss below the conditions needed for two shells $a,\,b$ to be open.
The exchange term in Hamiltonian \ref{Equation:H_0} promotes that $S_T$ is maximized
if the two shells are open, because $V_{ab}^0$ is always positive.

The low-energy Hamiltonian  can be diagonalized analytically for $O=2$ or $3$, but becomes analytically unmanageable
for larger $O$ because of the exponential size growth of its Fock space. However, the mechanisms
leading to open shell and high-spin behavior are captured already for $O=2$ and $O=3$. These mechanisms
are discussed in the following two sections. The conditions that will be found below 
for the existence of open-shell high-spin states are equivalent to similar conditions that have been established 
for the Hubbard  model \cite{Pieri96,Tasaki98}. Those results are also akin to the mesoscopic
Stoner instabilities predicted to exist in open quantum dots \cite{Kurland00,Burmistrov20}.


\subsection{Two shell model}
The two-shell model illustrates how more than one shell can be open at the same time. The Hamiltonian 
is
\begin{eqnarray}
\label{Equation:H2}
\hat{H}&=&\,\epsilon_1\,\hat{n}_1+{\cal U}\,\hat{n}_{1\uparrow}\,\hat{n}_{1\downarrow}
+\epsilon_2\,\hat{n}_2+{\cal U}\,\hat{n}_{2\uparrow}\,\hat{n}_{2\downarrow}\\
&&+V_{12}\left(\,\frac{\hat{n}_1\,\hat{n}_2}{2}-2\,\hat{\bf{S}}_1\cdot\hat{\bf{S}}_2
+\hat{\Delta}_1^\dagger\,\hat{\Delta}_2+\hat{\Delta}_2^\dagger\,\hat{\Delta}_1\right)\nonumber
\end{eqnarray}
where we have dropped the super-indices {\it MF} and $0$, and where we assume that $\epsilon_1<\epsilon_2$.
In addition, $V_{12}$ can be equal to
${\cal V}$ or ${\cal W}$, depending on wether $1$ and $2$ belong to the same band or not.
The Fock space has five subspaces
subspaces labeled by $N=0,\,1,\,2,\,3,\,4$. The highest spin subspace $S_T=1$ is achieved
at half-filling $N=2$. The ground state for $N=1$ corresponds to the open-shell doublet
$|1,0,0,0)$, $|0,1,0,0)$ with energy $E_{1,1,1/2,\pm1/2}=\epsilon_1$. The subspace $N=2$
has two closed-shell singlets with energies and wavefunctions
\begin{eqnarray}
    E_{2,1-2,0,0}&=&\epsilon_1+\epsilon_2+{\cal U}\pm\sqrt{(\Delta\epsilon_{21})^2+V_{12}^2}\\
    \ket{2,1-2,0,0}&=&A_+ |1,1,0,0)\pm A_- |0,0,1,1)\\
    A_{\pm}&=&\frac{1}{\sqrt{2}}\,\left(1\mp \frac{\Delta\epsilon_{21}}{\sqrt{(\Delta\epsilon_{21})^2+V_{12}^2}}\right)^{1/2}
\end{eqnarray}
with $\Delta\epsilon_{21}=\epsilon_2-\epsilon_1$. The singlet $\ket{2,2,0,0}$ has a higher weight in the single-particle
state $|1,1,0,0)$ and viceversa.
The subspace also contains an open-shell triplet 
\begin{eqnarray}
    E_{2,1,1,1-3}&=&\epsilon_{1}+\epsilon_{2}\\
    \ket{2,1,1,1}&=&|1,0,1,0)\\
    \ket{2,1,1,0}&=&\frac{|1,0,0,1)+|0,1,1,0)}{\sqrt{2}}\\
    \ket{2,1,1,-1}&=&|0,1,0,1)
\end{eqnarray}
and an open-shell singlet
\begin{eqnarray}
     E_{2,3,0,0}&=&\epsilon_{1}+\epsilon_{2}+2 \,V_{12}\\
    \ket{2,3,0,0}&=&\frac{|1,0,0,1)-|0,1,1,0)}{\sqrt{2}}
\end{eqnarray}
The open shell singlet-triplet energy difference defines the exchange energy 
\begin{equation}
    J=\frac{1}{2}\left(E_\mathrm{singlet}-E_\mathrm{triplet}\right)=V_{12}
\end{equation} 
where the convention is that $J>0$ means ferromagnetic alignment. 

\begin{figure}[ht]  \centering
  \includegraphics[width=0.5\textwidth]{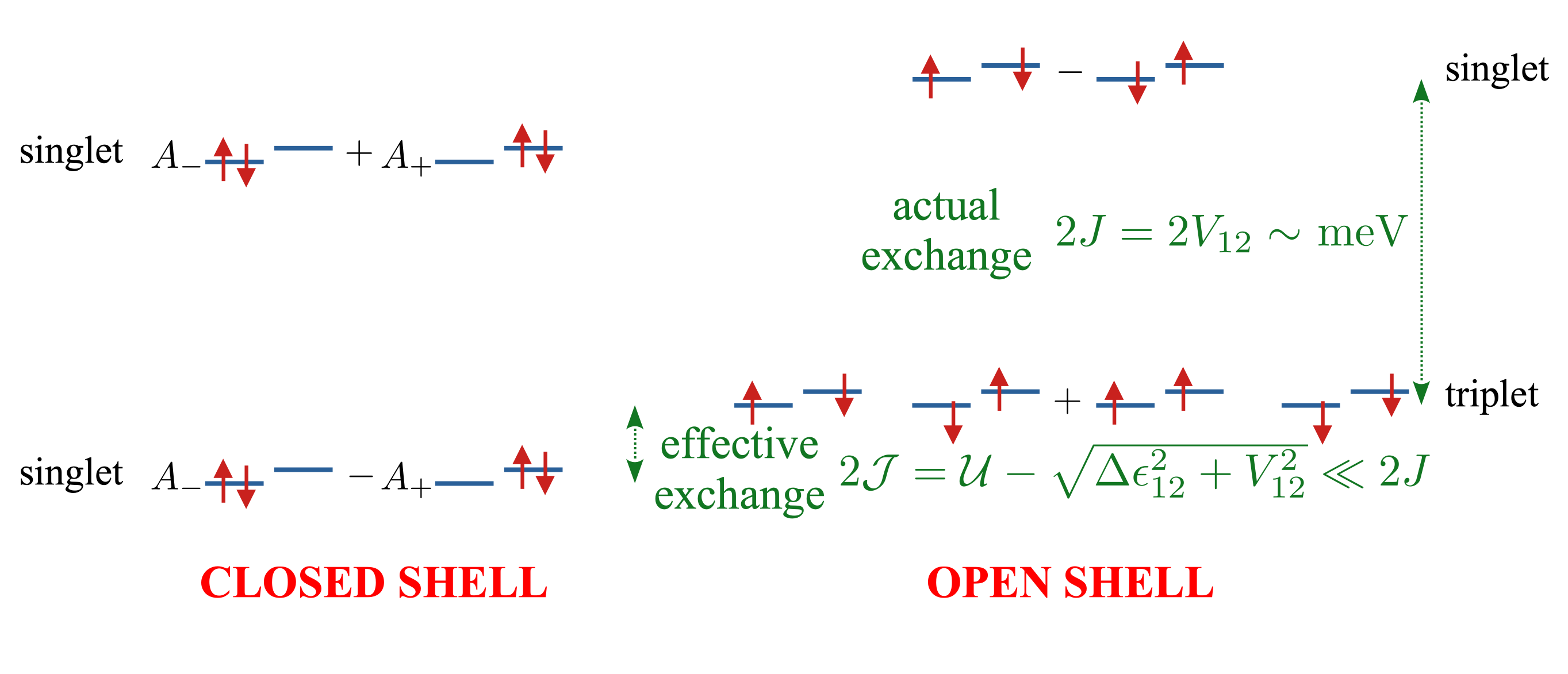}
  \caption{Illustration of the level ordering for the $N=2$ subspace of the two shell model.}
  \label{Figure:sketch}
\end{figure}

On the other hand, the ground state is either the closed shell singlet $\ket{2,2,0,0}$ or the 
open shell triplet $\ket{2,1,1,1-3}$ depending on whether the {\it effective} exchange constant 
\begin{eqnarray}
    {\cal J}&=&\frac{1}{2}\left(E'_\mathrm{singlet}-E_\mathrm{triplet}\right)
    =\frac{1}{2}\left({\cal U}-\sqrt{(\Delta\epsilon_{21})^2+V_{12}^2}\right)\nonumber\\
\end{eqnarray}
is smaller or larger than zero, respectively. Notice that if the two
shells are open at the same time, then the effective exchange is positive (e.g.: ferromagnetic) and 
the ground state has high spin $S=1$. The full level 
ordering  in the $N=2$ subspace is illustrated in Fig. \ref{Figure:sketch}.

\subsection{Three shell model}
The three-shell model  describes how a higher spin $S=3/2$ 
can emerge in an armchair ribbon. The Hamiltonian is
\begin{eqnarray}
\hat{H}&=&\sum_{i=1,2,3}\,
\left(\epsilon_i\,\hat{n}_i+{\cal U}\,\hat{n}_{i\uparrow}\,\hat{n}_{i\downarrow}\right)\\
&&+V_{12}\left(\,\frac{\hat{n}_1\,\hat{n}_2}{2}-2\,\hat{\bf{S}}_1\cdot\hat{\bf{S}}_2
+\hat{\Delta}_1^\dagger\,\hat{\Delta}_2+\hat{\Delta}_2^\dagger\,\hat{\Delta}_1\right)\nonumber\\
&&
+V_{13}\left(\,\frac{\hat{n}_1\,\hat{n}_3}{2}-2\,\hat{\bf{S}}_1\cdot\hat{\bf{S}}_3
+\hat{\Delta}_1^\dagger\,\hat{\Delta}_3+\hat{\Delta}_3^\dagger\,\hat{\Delta}_1\right)\nonumber\\
&&+V_{23}\left(\,\frac{\hat{n}_2\,\hat{n}_3}{2}-2\,\hat{\bf{S}}_2\cdot\hat{\bf{S}}_3
+\hat{\Delta}_2^\dagger\,\hat{\Delta}_3+\hat{\Delta}_3^\dagger\,\hat{\Delta}_2\right)\nonumber
\end{eqnarray}
with $\epsilon_1<\epsilon_2<\epsilon_3$. 
In addition, $V_{a b}$ can be equal to
${\cal V}$ or ${\cal W}$, depending on wether $a$ and $b$ belong to the same band or not.
The Hamiltonian is diagonalized in detail in the appendix. The Fock space has seven 
subspaces labeled by $N=0,\,1,\,\dots,\,6$. The highest spin subspace $S_T=3/2$ is achieved
at half-filling $N=3$.

We discuss now the conditions for {\it shell} opening and closing for the $N=2$ subspace.
The subspace contains three triplet ($S=1$) and six singlet eigen-states. We analyse the 
limit $\epsilon_3\gg\epsilon_{1,2}$ to provide a simpler discussion. The two relevant 
shells are then $i=1,\,2$. We find that the lowest singlet corresponds always to a close shell 
situation, while the triplets correspond to open shells. The energy difference between the
lowest-lying triplet and singlet states then tells whether the two shells are open or closed, and also 
defines the effective exchange constant 
\begin{equation}
    \label{Equation:twoshell}
    {\cal J}=\frac{1}{2}\left(E_\mathrm{singlet}-E_\mathrm{triplet}\right)={\cal U}-\sqrt{(\Delta \epsilon_{21})^2+V_{12}^2}
\end{equation}
In other words, if the two shells are open, then the effective exchange is ferromagnetic ${\cal J}< 0$, and vice versa.

The sub-space $N=3$ corresponds to the half-filled model. It contains 20 states that are split 
into 8 doublets with $S=1/2$ and a quadruplet with $S=3/2$. Six doublets 
contain one fully occupied, one half-occupied and one empty shell. The remaining
two have three open shells but their eigen-energies are too large.
The quadruplet corresponds to the three shells being open at the same time. The quadruplet is the ground state if the conditions 
\begin{equation}
\label{Equation:threeshell}
{\cal U}+\frac{V_{ab}+V_{cb}}{2}>\sqrt{\left(\Delta \epsilon_{ac}+\frac{V_{ab}-V_{cb}}{2}\right)^2+(V_{ac})^2}
\end{equation}
for the three possible configurations $a\neq b\neq c=1,\,2,\,3$ are fulfilled at the same time.
The tightest of these conditions defines the energy difference between the lowest
quadruplet and the lowest doublet, and the effective exchange constant ${\cal J}$. So if the three shells are open,
then ${\cal J}$ is ferromagnetic; otherwise one shell is closed, another is empty and a third
is half-filled.
In addition, several of the doublet states contain one full or one empty shell. The model 
could in these cases be reduced to a simpler two-shell model. Furthermore, this feature 
illustrates the mechanism by which a model containing a large number of shells folds into a 
simpler effective model with fewer shells. 

The discussion above can be extended to larger $O$-shell models to determine the conditions for
the existence of spin-$O/2$ states, as well as the probability of finding them in a given 
GNR of size $(M_x,\,M_y)$. The exponential growth of the Fock space prevents however the 
analytical determination of these conditions and probabilities. The whole discussion, however 
sheds strong indications that states with spin as large as 3 or even 4 may be found in GNR.


\subsection{Example: a N=9 armchair ribbon}
To provide a more explicit example, we discuss the armchair GNR with $N=9$, $M_x=80$ introduced in 
Fig. \ref{Figure:bands} in view of the three shell model, 
although this discussion applies to any ribbon with similar widths and lengths, as shown in 
Table \ref{Table:percentages}.
Our estimates indicate that this free-standing ribbon has a Coulomb ${\cal U}\approx 6.8$ meV.
Shell pairs or triples with energy differences smaller than $6.8$ meV are candidates to 
show open shell high-spin behavior, according to Equs. \ref{Equation:twoshell} and 
\ref{Equation:threeshell}. The two panels in Fig. \ref{Figure:addition_energies} show that a 
significant fraction of shell pairs may fulfill these conditions. We have determined that the
fraction of shells fulfilling the condition \ref{Equation:twoshell} is $P_{S=1}\approx 29$ per cent. 
Similarly, we have also found that a fraction $P_{S=3/2}\sim 17$ per cent of the 
shells fulfill the three conditions in Eq. \ref{Equation:threeshell} for the existence of $S=3/2$ 
states in this ribbon. We have finally computed these two fractions for GNR of different sizes 
to estimate the likelihood to find high-spin open-shell behavior. 
The results are summarized in 
Table \ref{Table:percentages} and demonstrate that GNRs should contain a non-negligible
number of open shells. The table indicates that the chances to display double and triple 
open shell behavior increase, but saturate, with the length and width of the ribbon. 

Graphene ribbons are
however experimentally addressed when they are deposited on a substrate. Furthermore, in electrical
experiments, the ribbons are contacted at the two sides by metallic electrodes. To provide a 
phenomenological understanding of these substrate and contact effects, we have included an
effective dielectric constant $\varepsilon$ in our calculations, so that $U^{eff}=U^0/\varepsilon$. 
Table \ref{Table:percentages} 
shows in parentheses the percentages found when $\varepsilon=2$. Dielectric screening reduces
the fraction of open shells because it decreases the size of the effective Coulomb matrix 
elements.

\setlength{\tabcolsep}{4pt}
\begin{table}[ht]
    \centering
    \begin{tabular}{|c|c|c|c|c|c|}
    \hline
   $M_x$ & 10 & 100 & 500 \\
   \hline
   $M_y$ &                  &                    &      \\ 
   \hline
    4    & 22 (12) /  7 (0) & 28 (14) / 15 (4)   & 28 (15) / 17 (6) \\
    \hline
    5    & 29 (14) / 13 (4) & 30 (16) / 19 (6)   & 31 (17) / 19 (7) \\
    \hline
    9    & 34 (20) / 19 (4) & 36 (17) / 24  (8)  & 36 (21) / 25 (9) \\
    \hline
   16    & 31 (18) / 19 (3) & 39 (21) / 27  (8)  & 40 (22) / 29 (10) \\
   \hline
   26    & 34 (20) / 24 (8) & 41 (23) / 30  (11) & 42 (24) / 31 (12) \\
   \hline
   36    & 38 (19) / 24 (7) & 41 (23) / 30  (11) & 42 (24) / 32 (12) \\
   \hline
    \end{tabular}
    \caption{Percentage of states fulfilling the conditions for $S=1$ / $S=3/2$ for the half-filled 
    two- / three-shell models in a doped armchair GNR of length $M_x$ and width $M_y$. Numbers inside the
    parentheses indicate the percentages when an effective dielectric constant $\varepsilon=2$ 
    is taken into account.}
    \label{Table:percentages}
\end{table}


\section{Electronic transport through doped graphene nanoribbons}


Graphene ribbons may have use in different technologies, and we single out here their possible
use as channel components in nanoelectronics \cite{Zhang23a,Zhang23b,Niu23,Zhang26}.
Electronic transport through a ribbon connected to two electrodes $A=L,\,R$ is governed by the transfer Hamiltonian
\begin{eqnarray}
    \hat{H}_t&=&\sum_{A}\,\sum_{k,i,\sigma}\, 
    \left(\,{\cal V}_{A,k,i,\sigma}\,\hat{r}_{i\sigma}^\dagger\,\hat{c}_{A,k\sigma}\,+h.c.\right)\\
    &=&
    \sum_{A}\,\sum_{k,\sigma,\lambda,\lambda'}
    \left(\,
    {\cal V}_{A,k\sigma;N\lambda\rightarrow N+1\lambda'}\,\hat{\Gamma}_{N+1\lambda'}^\dagger\,\hat{\Gamma}_{N\lambda}\,
    \hat{c}_{A,k\sigma}+h.c.\right)\nonumber
\end{eqnarray}
where $\hat{c}_{A,k\sigma}^\dagger$ creates an electron with spin $\sigma$ and wave-vector $k$ in electrode $A$ and ${\cal V}_{A,k,i\sigma}$ denotes a 
single-particle transfer matrix element. The matrix elements are written in the many-body
eigen-state basis as follows
\begin{eqnarray}
{\cal V}_{A,k\sigma;N\lambda\rightarrow N+1 \lambda'}&=&
\sum_i\,{\cal V}_{A,k,i\sigma}\,M_{\lambda'\lambda}^N(i,\sigma)\\
M_{\lambda'\lambda}^N(i,\sigma)&=&\braket{N+1\lambda'\,|\hat{r}_{i\sigma}^\dagger|\,N\lambda}\\
&=&
({\cal R}^{N+1})^\dagger_{\lambda'n'}\,(n'|\hat{r}_{i\sigma}^\dagger| n) \,{\cal R}^N_{n \lambda}\nonumber
\end{eqnarray}
where we have made use of Eq. \ref{Equation:rotation}. The 
transfer rates in the Single Electron Transfer regime (SET) are 
\begin{eqnarray}
\label{Equation:rates}
 \gamma_{A,N\lambda\rightarrow N+1\lambda'}&=&\frac{2\,\pi}{\hbar}\,g_{A}(\mu_A)\,f(E_{N+1\lambda'}-E_{N\lambda}-\mu_A)\times\nonumber\\
 &&\times\sum_{i\sigma}\,\,|{\cal 
V}_{A,k_F,i,\sigma}|^2\,|M^N_{\lambda'\lambda}(i,\sigma)|^2   \\
 \gamma_{A,N\lambda\rightarrow N-1\lambda'}&=&\frac{2\,\pi}{\hbar}\,g_{A}(\mu_A)\,(1-f(E_{N\lambda}-E_{N-1\lambda'}-\mu_A))\times\nonumber\\
  &&\times\sum_{i\sigma}\,\,|{\cal 
  V}_{A,k_F,i,\sigma}|^2\,|M^{N-1}_{\lambda'\lambda}(i,\sigma)|^2 
\end{eqnarray}
 where $g_{A}(\mu_A)$ is the density of states at the chemical 
 potential $\mu_A$ of the $A$ electrode.

The above expressions for electron transfer together with the conditions
for open shell behavior found in this article lead to a clear mechanism for spin and shell
blockade in the transport response of nanoribbons. The mechanism for shell blockade in the 
SET regime is illustrated with the following example: the transition matrix element 
$(1,1,0,0,0,0|\hat{r}_{i\sigma}^\dagger|0,0,1,0,0,0)$ is always zero 
because  the transition needs two electrons to change their single-particle state 
simultaneously. Similarly, the origin of the 
mechanism for spin-blockade in the SET regime lies in spin-conservation. For 
example, the transition matrix element
$\braket{N=2,R,S=0,S_z |\hat{r}_{i\sigma}^\dagger| N=3,R',S=3/2,S_z'}$ is always zero
because a single electron can raise the spin by only $1/2$.

The explicit expressions for the three-shell model are
\begin{eqnarray}
\label{Equation:rates}
 \gamma_{A,N\lambda\rightarrow N+1\lambda'}&=&
 \gamma_{A,1}\,|M^N_{\lambda'\lambda}(1)|^2+
 \gamma_{A,2}\,|M^N_{\lambda'\lambda}(2)|^2\nonumber\\
 &&+\gamma_{A,3}\,|M^N_{\lambda'\lambda}(3)|^2
\end{eqnarray}
where we have defined 
$\gamma_{A,i}=\frac{2\pi}{\hbar}\,g_A(\mu_A)\,|{\cal V}_{A,k_F,i}|^2$ and 
$|M^N_{\lambda'\lambda}(i)|^2=|M^N_{\lambda'\lambda}(i,\uparrow)|^2+|M^N_{\lambda'\lambda}(i,\downarrow)|^2$.


\section{Conclusions}


In conclusion, we have addressed the electronic behavior of free-standing doped GNR
using an extended Hubbard model. We have performed an exact mapping of the model to a 
Kanamori Hamiltonian. We have reduced the range of the Hubbard parameters, and showed
that an onsite Hubbard model with $U\approx 4.9$ eV fits well the low-lying mean field
solutions to DFT simulations of ribbons of different widths and lengths. We have determined
from the onsite model, analytical equations for the Coulomb matrix elements. We have found 
that those matrix elements follow a remarkable
scaling law, and that they collapse into just three for each ribbon. We have then
introduced an effective low-energy Hamiltonian to address the many-body behavior of 
ribbons where the chemical potential lies inside the bulk bands. The model allows us to
unveil a robust mechanism that stabilizes open-shell high-spin states. We have estimated the
percentage of cases where open-shell high-spin states are expected to exist. These percentages
increase but saturate with the ribbon lengths and widths and decrease when an 
effective dielectric constant is included in the calculations.


\begin{acknowledgments}


JF would like to acknowledge very useful discussions with Malte R\"{o}ssner and technical assistance from
Alvaro Correal. JF and AGF have been supported by MCIN/AEI/10.13039/501100011033/FEDER, by UE and by 
SEKUENS through the projects PID2022-137078NB-100, TRILMAX 101159646, and 
UONANO IDE/2024/000678. At Delft, the work was supported by a HORIZON-EIC-2022-PATHFINDEROPEN 
project (ATYPICAL, project number 101099098). 
\end{acknowledgments}


\appendix


\section{Diagonalization of the three shell model}

The Hamiltonian for the three-shell model is
\begin{eqnarray}
\hat{H}&=&\sum_{i=1,2,3}\,
\left(\epsilon_i\,\hat{n}_i+{\cal U}\,\hat{n}_{i\uparrow}\,\hat{n}_{i\downarrow}\right)\\
&&+V_{12}\left(\,\frac{\hat{n}_1\,\hat{n}_2}{2}-2\,\hat{\bf{S}}_1\cdot\hat{\bf{S}}_2
+\hat{\Delta}_1^\dagger\,\hat{\Delta}_2+\hat{\Delta}_2^\dagger\,\hat{\Delta}_1\right)\nonumber\\
&&
+V_{13}\left(\,\frac{\hat{n}_1\,\hat{n}_3}{2}-2\,\hat{\bf{S}}_1\cdot\hat{\bf{S}}_3
+\hat{\Delta}_1^\dagger\,\hat{\Delta}_3+\hat{\Delta}_3^\dagger\,\hat{\Delta}_1\right)\nonumber\\
&&+V_{23}\left(\,\frac{\hat{n}_2\,\hat{n}_3}{2}-2\,\hat{\bf{S}}_2\cdot\hat{\bf{S}}_3
+\hat{\Delta}_2^\dagger\,\hat{\Delta}_3+\hat{\Delta}_3^\dagger\,\hat{\Delta}_2\right)\nonumber
\end{eqnarray}
The states are ordered so that 
$\epsilon_1<\epsilon_2<\epsilon_3$. The Fock space ${\cal F}$ has dimensions 
$4^3=64\,$ and is spanned by the single-particle basis 
$\{ |\,n_{1\uparrow},\,n_{1\downarrow},n_{2\uparrow},\,n_{2\downarrow},
n_{3\uparrow},\,n_{3\downarrow}) \}$. 
The Hamiltonian commutes with the total number operator  $\hat{N}$. $\hat{N}$ has eigen-values 
$N =0,\,1,\,...,\,6$, so that ${\cal F}$ is split into seven subspaces labeled by $N$. The
Hamiltonian also commutes with total spin operator $\hat{S}^2$, so that each of those subspaces is further 
split into subspaces labeled by $S$. However, $\hat{H}$, $\hat{N}$ and $\hat{S}$ do not constitute
a complete set of commuting observables because there exist for some of the sub-spaces, more than 
one same-$S$ multiplet. We label these different subspaces with an additional integer number $R$ whenever
they exist. All in all, we use the following notation to describe the Hamiltonian eigen-states: 
$\ket{N,\,R,\,S,\,S^z}$. We describe below the subspaces $N=0,\,1,\,2$ and 3, their eigen-states and 
their energies. Sub-spaces $N=4,\,5$ and 6 are easily deduced from the electron-hole symmetry of
the Hamiltonian. The highest spin subspace occurs at half-filling, $N=3$, and
corresponds to a quadruplet with $S=3/2$. We denote $\Delta\epsilon_{ab}=\epsilon_a-\epsilon_b$.

\setlength{\tabcolsep}{10pt}
\begin{itemize}
\item{Subspace N=0.} This trivial subspace is one-dimensional and contains only the vacuum state
\newline\newline
\begin{tabular}{lll}
$\ket{0,0,0}$ &= $\ket{000000}$ & $E = 0$
\end{tabular}
\item{Subspace N=1.} This subspace has dimensions equal to 6 and is made up of the following three doublets
\newline\newline
\begin{tabular}{lll}
$\ket{1,1,1/2,1/2}$  &= $\ket{100000}$& $E \,=\,\epsilon_1$ \\
$\ket{1,1,1/2,-1/2}$ &= $\ket{010000}$& $E \,=\,\epsilon_1$ \\\\
$\ket{1,2,1/2,1/2}$  &= $\ket{001000}$& $E \,=\,\epsilon_2$ \\
$\ket{1,2,1/2,-1/2}$ &= $\ket{000100}$& $E \,=\,\epsilon_2$ \\\\
$\ket{1,3,1/2,1/2}$  &= $\ket{000010}$& $E \,=\,\epsilon_3$ \\
$\ket{1,3,1/2,-1/2}$ &= $\ket{000001}$& $E \,=\,\epsilon_3$ \\
\end{tabular}

\item{Subspace N=2.} This subspace has dimensions equal to 15 and is made up of six singlets and three triplets.
The Hamiltonian for the basis states $\ket{1,1,0,0,0,0}$, $\ket{0,0,1,1,0,0}$ and $\ket{0,0,0,0,1,1}$ is
\begin{eqnarray}
    \left(\begin{array}{ccc}
    2\epsilon_1+{\cal U}&V_{12}&V_{13}\\
    V_{12}&2\,\epsilon_2+{\cal U}&V_{23}\\
    V_{13}&V_{23}&2\,\epsilon_3+{\cal U}
    \end{array}\right)
\end{eqnarray}
Solving the eigen-value equation yields a complicated cubic equation, whose explicit solution delivers
three eigen-energies. The corresponding three eigenstates are singlets of the form
\begin{eqnarray*}
    \ket{2,\,R,\,0,\,0}&=&A_{R1}\ket{1,1,0,0,0,0}+A_{R2}\ket{0,0,1,1,0,0}\\&&
    +A_{R3}\ket{0,0,0,0,1,1}
\end{eqnarray*}
where $R=1,\,2,\,3$. 
The remaining three singlet eigen-states and eigen-energies  are
\newline\newline
\begin{tabular}{lll}
$\ket{2,4,0,0}$  &= $\frac{\ket{100100}-\ket{011000}}{\sqrt{2}}$& $E \,=\,\epsilon_1+\epsilon_2+2\,V_{12}$ \\
$\ket{2,5,0,0}$  &= $\frac{\ket{100001}-\ket{010010}}{\sqrt{2}}$& $E \,=\,\epsilon_1+\epsilon_3+2\,V_{13}$ \\
$\ket{2,6,0,0}$  &= $\frac{\ket{001001}-\ket{000110}}{\sqrt{2}}$& $E \,=\,\epsilon_2+\epsilon_3+2\,V_{23}$ 
\end{tabular}
\newline \newline
The three triplets are
\newline\newline
\begin{tabular}{lll}
$\ket{2,1,1,1}$  &= $\ket{101000}$& $E \,=\,\epsilon_1+\epsilon_2$ \\
$\ket{2,1,1,0}$  &= $\frac{\ket{100100}+\ket{011000}}{\sqrt{2}}$& $E \,=\,\epsilon_1+\epsilon_2$ \\
$\ket{2,1,1,-1}$ &= $\ket{010100}$& $E \,=\,\epsilon_1+\epsilon_2$\\\\
$\ket{2,2,1,1}$  &= $\ket{100010}$& $E \,=\,\epsilon_1+\epsilon_3$ \\
$\ket{2,2,1,0}$  &= $\frac{\ket{100001}+\ket{010010}}{\sqrt{2}}$& $E \,=\,\epsilon_1+\epsilon_3$ \\
$\ket{2,2,1,-1}$ &= $\ket{010001}$& $E \,=\,\epsilon_1+\epsilon_3$\\\\
$\ket{2,3,1,1}$  &= $\ket{001010}$& $E \,=\,\epsilon_2+\epsilon_3$ \\
$\ket{2,3,1,0}$  &= $\frac{\ket{000110}+\ket{001001}}{\sqrt{2}}$& $E \,=\,\epsilon_2+\epsilon_3$ \\
$\ket{2,3,1,-1}$ &= $\ket{000101}$& $E \,=\,\epsilon_2+\epsilon_3$
\end{tabular}
\newline\newline
The lowest-lying states are the closed shell singlet $\ket{2,\,1,\,0,\,0}$ and the open shell 
triplet $\ket{2,1,1,S^z}$.
The condition for the ground state to be a triplet with spin $S=1$ can be found only numerically.

We analyse here two situations in more detail. The first highlights how effective 
a ferromagnetic exchange energy arises, and how a third shell may be or may be not enter in the
calculatiions. The second highlight how a shell becomes closed.
\begin{enumerate}
    \item{Limit $\epsilon_3\gg\epsilon_2$.}\newline
    Here, the third shell is completely empty. The problem thus reduces to a simpler 
    half-filled two-shell model. The two singlet states simplify to 
\begin{eqnarray*}
    \ket{2,\,R,\,0,\,0}=A_{R1}\ket{1,1,0,0}+A_{R2}\ket{0,0,1,1}
\end{eqnarray*}
where $R=1,\,2$, with eigen-energies 
\begin{eqnarray*}
    E&=&\epsilon_1+\epsilon_2+{\cal U}\pm\sqrt(\Delta{\epsilon_{21})^2+V_{12}^2}
\end{eqnarray*}
The energy difference between the singlet and the triplet defines an effective
exchange coupling 
\begin{equation}
{\cal J}={\cal U}-\sqrt(\Delta{\epsilon_{21})^2+V_{12}^2}
\end{equation}
The triplet state $\ket{2,\,1,\,1,\,S^z}$ becomes the ground state if ${\cal J}>0$,
e.g.: if the effective exchange coupling becomes ferromagnetic.

\item{Limit $\epsilon_1\ll\epsilon_2$.}\newline
This limit corresponds to the closing on the lowest shell. Again the equations simplify so that
the singlet states are
\begin{eqnarray*}
    \ket{2,\,R,\,0,\,0}=A_{R1}\ket{1,1,0,0}+A_{R2}\ket{0,0,1,1}
\end{eqnarray*}
where $R=2,\,3$, and with eigen-energies 
\begin{eqnarray*}
    E&=&\epsilon_2+\epsilon_3+{\cal U}\pm\sqrt{(\Delta\epsilon_{32})^2+V_{23}^2}
\end{eqnarray*}
\end{enumerate}

\item{Subspace N=3.} This subspace has dimensions equal to 20 and is made up of eight doublets and a quadruplet. 
We list first all the basis 
states
   \begin{eqnarray*}
    \begin{split}
        \ket{A,S^z=1/2}&=\ket{111000}&\hspace{20pt}E_A&=2\,\epsilon_1+\epsilon_2+{\cal U}+V_{12}\\
        \ket{B,S^z=-1/2}&=\ket{110100}&\hspace{20pt}E_B&=E_A\\ 
        \ket{C,S^z=1/2}&=\ket{001011}&\hspace{20pt}E_C&=\epsilon_2+2\,\epsilon_3+{\cal U}+V_{23}\\
        \ket{D,S^z=-1/2}&=\ket{000111}&\hspace{20pt}E_D&=E_C\\
        \ket{E,S^Z=1/2}&=\ket{110010}&\hspace{20pt}E_E&=2\,\epsilon_1+\epsilon_3+{\cal U}+V_{13} \\ 
        \ket{F,S^Z=-1/2}&=\ket{110001}&\hspace{20pt}E_F&=E_E\\ 
        \ket{G,S^z=1/2}&=\ket{001110}&\hspace{20pt}E_G&=2\,\epsilon_2+\epsilon_3+{\cal U}+V_{23}\\ 
        \ket{H,S^z=-1/2}&=\ket{001101}&\hspace{20pt}E_H&=E_G\\
        \ket{I,S^Z=1/2}&=\ket{101100}&\hspace{20pt} E_I&=\epsilon_1+2\,\epsilon_2+{\cal U}+V_{12} \\
        \ket{J,S^Z=-1/2}&=\ket{011100}&\hspace{20pt}E_J&=E_I\\ 
        \ket{K,S^z=1/2}&=\ket{100011}&\hspace{20pt}E_K&=\epsilon_1+2\,\epsilon_3+{\cal U}+V_{13} \\ 
        \ket{L,S^z=-1/2}&=\ket{010011}&\hspace{20pt}E_L&=E_K\\
        \ket{M,S^z=1/2} &= \ket{101001}&\hspace{20pt}E_M&=\epsilon_a+V_{13}+V_{23}\\
        \ket{N,S^z=1/2} &= \ket{100110}&\hspace{20pt}E_N&=\epsilon_a+V_{12}+V_{23}\\
        \ket{O,S^z=1/2} &= \ket{011010}&\hspace{20pt}E_O&=\epsilon_a+V_{12}+V_{13}\\
        \ket{P,S^z=-1/2} &= \ket{010110}&\hspace{20pt}E_P&=\epsilon_a+V_{13}+V_{23}\\
        \ket{Q,S^z=-1/2} &= \ket{011001}&\hspace{20pt}E_Q&=\epsilon_a+V_{12}+V_{23}\\
        \ket{R,S^z=-1/2} &= \ket{100101}&\hspace{20pt}E_R&=\epsilon_a+V_{12}+V_{13}
    \end{split}
\end{eqnarray*} 
where $\epsilon_a=\epsilon_1+\epsilon_2+\epsilon_3$ and 
$\epsilon_b=\epsilon_a+\left(V_{12}+V_{13}+V_{23}\right)$. 
The first set has six doublets, all of them having one fully occupied, one 
half-occupied and one empty shell. The first two doublets in this set are made of the 
states $\ket{A}$ and $\ket{C}$ for spin $S^Z=+1/2$ and $\ket{B}$ and $\ket{D}$ 
for spin $S^Z=-1/2$. They have inter-state couplings $V_{AC}=V_{BD}=V_{13}$. Both Hamiltonians 
are 
identical, so we write just the first one:
\begin{eqnarray}
    \left(\begin{array}{cc}
    E_A& V_{AB}\\V_{AB}&E_B
    \end{array}\right)
\end{eqnarray}

The doublets' eigen-states and eigen-energies $\ket{N,L,S,S^z}$ are
\begin{eqnarray*}
    \begin{split}
    \ket{3,1,1/2,1/2} &= u_+\,\ket{A} - u_-\,\ket{C}\\
    \ket{3,1,1/2,-1/2}&= u_+\,\ket{B} - u_-\,\ket{D}\\
    \epsilon_{3,1,1/2,\pm1/2}&=\frac{E_A+E_C}{2}-\sqrt{\left(\frac{E_C-E_A}{2}\right)^2+V_{AC}^2}\\
    &=\epsilon_a+{\cal U}+\frac{V_{12}+V_{23}}{2}\\&-
    \sqrt{\left(\Delta\epsilon_{31}+\frac{V_{23}-V_{12}}{2}\right)^2+V_{13}^2}\\
    \ket{3,2,1/2,1/2} &= u_-\,\ket{A} + u_+-\,\ket{C}\\
    \ket{3,2,1/2,-1/2}&= u_-\,\ket{B} + u_+-\,\ket{D}\\
    \epsilon_{3,2,1/2,\pm1/2}&=\frac{E_A+E_C}{2}+\sqrt{\left(\frac{E_C-E_A}{2}\right)^2+V_{AC}^2}\\
    &=\epsilon_a+{\cal U}+\frac{V_{12}+V_{23}}{2}\\&+
    \sqrt{\left(\Delta\epsilon_{31}+\frac{V_{23}-V_{12}}{2}\right)^2+V_{13}^2}\\
    \end{split}
\end{eqnarray*}
The next two doublets are made of the states $\ket{E}$ and $\ket{G}$ for spin $S^Z=+1/2$ and $\ket{F}$ and $\ket{H}$ 
for spin $S^Z=-1/2$. Their inter-state couplings $V_{EG}=V_{FH}=V_{12}$.  The Hamiltonian, eigen-states and
eigen-energies are the same as above upon the replacement $A\rightarrow E$, $B\rightarrow F$, $C\rightarrow G$ 
and $D\rightarrow H$, and the doublets are labeled now as $\ket{3,3,1/2,\pm1/2}$ and $\ket{3,4,1/2,1/2}$. Their eigen-energies are
\begin{eqnarray*}
    E_{3,3,1/2,\pm1/2}&=&\epsilon_a+{\cal U}+\frac{V_{13}+V_{23}}{2}\\&&-
    \sqrt{\left(\Delta\epsilon_{21}+\frac{V_{23}-V_{13}}{2}\right)^2+V_{12}^2}\\
    E_{3,4,1/2,\pm1/2}&=&\epsilon_a+{\cal U}+\frac{V_{13}+V_{23}}{2}\\&&+
    \sqrt{\left(\Delta\epsilon_{21}+\frac{V_{23}-V_{13}}{2}\right)^2+V_{12}^2}
\end{eqnarray*}

The last two doublets in this set are made of the states $\ket{I}$ and $\ket{K}$ for spin $S^Z=+1/2$ and $\ket{J}$ 
and $\ket{L}$ for spin $S^Z=-1/2$. Their inter-state couplings are $V_{IK}=V_{JL}=V_{23}$. The Hamiltonian, 
eigen-states and eigen-energies are again the same as above upon the replacement $A\rightarrow I$, $B\rightarrow J$, 
$C\rightarrow K$ and $D\rightarrow L$, and the doublets are labeled now as $\ket{3,5,1/2,\pm1/2}$ and $\ket{3,6,1/2,1/2}$. Their eigen-energies are
\begin{eqnarray*}
    E_{3,5,1/2,\pm1/2}&=&\epsilon_a+{\cal U}+\frac{V_{13}+V_{12}}{2}\\&&-
    \sqrt{\left(\Delta\epsilon_{32}+\frac{V_{13}-V_{12}}{2}\right)^2+V_{23}^2}\nonumber\\
    E_{3,6,1/2,\pm1/2}&=&\epsilon_a+{\cal U}+\frac{V_{13}+V_{12}}{2}\\&&+
    \sqrt{\left(\Delta\epsilon_{32}+\frac{V_{13}-V_{12}}{2}\right)^2+V_{23}^2}\nonumber
\end{eqnarray*}

The last two doublets have three open shells, 
\begin{widetext}
\begin{eqnarray*}
\ket{3,7,1/2,1/2} &=& \frac{1}{D_+^{1/2}}\,\left((V_{23}-V_{12}-R)\,\ket{M}+(V_{13}-V_{23}+R)\,\ket{N}+(V_{12}-V_{13})\,\ket{O}\right) \\
 \ket{3,7,1/2,-1/2} &=& \frac{1}{D_+^{1/2}}\,\left((V_{23}-V_{12}-R)\,\ket{P}+(V_{13}-V_{23}+R)\,\ket{Q}+(V_{12}-V_{13})\,\ket{R}\right) \\
 \ket{3,8,1/2,1/2} &=& \frac{1}{D_-^{1/2}}\,\left((V_{23}-V_{12}+R)\,\ket{M}+(V_{13}-V_{23}-R)\,\ket{N}+(V_{12}-V_{13})\,\ket{O}\right) \\
 \ket{3,8,1/2,-1/2} &=& \frac{1}{D_-^{1/2}}\,\left((V_{23}-V_{12}+R)\,\ket{P}+(V_{13}-V_{23}-R)\,\ket{Q}+(V_{12}-V_{13})\,\ket{R}\right) \\
R&=&\frac{1}{2}\,\left((V_{12}-V_{13})^2+(V_{12}-V_{23})^2+(V_{13}-V_{23})^2\right)^{1/2}\nonumber\\ 
D_{\pm}&=&2\,R^{1/2}\,\left(R\pm\frac{1}{2}\,(V_{12}+V_{13})-V_{23}\right)^{1/2}\nonumber
\end{eqnarray*}
\end{widetext}
Their energies are 
\begin{eqnarray*}
    \epsilon_{3,7,\pm1/2}&=&\epsilon_b-R\\
    \epsilon_{3,8,\pm1/2}&=&\epsilon_b+R
    \end{eqnarray*}
    
The quadruplet eigen-states are
\begin{eqnarray*}
\begin{split}
\ket{3,3/2,3/2}  &= \ket{1,0,1,0,1,0} \\
\ket{3,3/2,1/2}  &= \frac{1}{\sqrt{3}}\,(\ket{M}+\ket{N}+\ket{O}) \\
\ket{3,3/2,-1/2} &= \frac{1}{\sqrt{3}}\,(\ket{P}+\ket{Q}+\ket{R}) \\
\ket{3,3/2,-3/2} &= \ket{0,1,0,1,0,1} \\
\end{split}
\end{eqnarray*}
and their eigen-energy is $E=\epsilon_a$.

Inspecting the above expressions for the eigen-energies, we come to the conclusion that the 
two doublets having all open shells have too large eigen-energies.
The quadruplet thus competes with the first set of six doublets to be the 
ground state.
The ground state will be the quadruplet state provided that the following three conditions are met
together:
\begin{eqnarray}
    {\cal U}+\frac{V_{12}+V_{23}}{2}-
    \sqrt{\left(\Delta\epsilon_{31}+\frac{V_{23}-V_{12}}{2}\right)^2+V_{13}^2}> 0\nonumber\\\\
    {\cal U}+\frac{V_{13}+V_{23}}{2}-
    \sqrt{\left(\Delta\epsilon_{21}+\frac{V_{23}-V_{13}}{2}\right)^2+V_{12}^2}> 0\nonumber\\\\
    {\cal U}+\frac{V_{13}+V_{12}}{2}
    \sqrt{\left(\Delta\epsilon_{32}+\frac{V_{13}-V_{12}}{2}\right)^2+V_{23}^2}> 0\nonumber\\
\end{eqnarray}
\end{itemize}


\bibliography{model.bib}

\end{document}